\documentclass[aps,prd,preprint,superscriptaddress,showpacs,floatfix,nobibnotes,nofootinbib]{revtex4-1}
\usepackage[usenames, dvipsnames]{color}

\usepackage{float}

\usepackage{latexsym}
\usepackage{amsmath}
\usepackage{amssymb}
\usepackage{graphicx}
\usepackage{setspace}
\usepackage{longtable}
\usepackage{enumitem}
\usepackage[bottom]{footmisc}

\usepackage[normalem]{ulem}
\usepackage[usenames, dvipsnames]{color}

\usepackage{setspace}

\usepackage{bm}
\usepackage{epsfig}
\newcommand{\bea}{\begin{eqnarray}}
\newcommand{\eea}{\end{eqnarray}}

\newcommand{\nn}{\nonumber}

\begin{document}
\setlength{\baselineskip}{20pt}

\title{Physical characteristics of glasma from the earliest stage of relativistic heavy ion collisions}

\author{Margaret E. Carrington}
\affiliation{Department of Physics, Brandon University,
Brandon, Manitoba R7A 6A9, Canada}
\affiliation{Winnipeg Institute for Theoretical Physics, Winnipeg, Manitoba, Canada}

\author{Alina Czajka}
\affiliation{National Centre for Nuclear Research, ul. Pasteura 7,  PL-02-093  Warsaw, Poland}

\author{Stanis\l aw Mr\' owczy\' nski}
\affiliation{Institute of Physics, Jan Kochanowski University, ul. Uniwersytecka 7, PL-25-406 Kielce, Poland }
\affiliation{National Centre for Nuclear Research, ul. Pasteura 7,  PL-02-093  Warsaw, Poland}

\date{August 04, 2022}

\begin{abstract}

We present some analytic results that describe the gluon field, or glasma, that exists at very early times after a collision of relativistic heavy ions at proper time $\tau=0$. We use a Colour Glass Condensate approach, and perform an expansion in $\tau$. 
The full details of our calculational method are described in our previous paper \cite{Carrington:2020ssh}, 
where we have explained all of the steps that are necessary to obtain the energy-momentum tensor up to sixth order in $\tau$. 
In this paper we present an analysis of various physical quantities that can be obtained from this energy-momentum tensor. 
We show that the expansion to order $\tau^6$ can be trusted to about $\tau=0.05$ fm.
We calculate the transverse and longitudinal pressures and show that, for times small enough that the expansion converges, they move towards their equilibrium values of one third of the energy density.
We also study the spatial eccentricity of the glasma, and the Fourier coefficients of the azimuthal flow. 
Our results for the Fourier coefficients are larger than  expected, which contradicts the usual assumption that anisotropy is mostly generated during the hydrodynamic evolution of the plasma. 
We  find a significant correlation between the elliptic flow coefficient and the eccentricity, which indicates that the spatial asymmetry introduced by the initial geometry is effectively transmitted to the azimuthal distribution of the gluon momentum field, even at very early times. 
This result is interesting because correlations of this kind are  characteristic of the onset of hydrodynamic behaviour. 
Finally, we have calculated the angular momentum of the glasma and obtained results that are many orders of magnitude smaller than the  angular momentum of the initial system of colliding ions in a configuration with non-zero impact parameter. 
This indicates that most of the angular momentum carried by the valence quarks is not transmitted to the glasma. 
The result is significant because it contradicts the picture of a rapidly rotating initial glasma state that has been proposed by several authors, but agrees with the current lack of experimental evidence for a significant polarization effect of the hyperons and vector mesons produced in heavy ion collisions at the highest accessible energies.

\end{abstract}

\maketitle

\section{Introduction}

We use a formulation of the Colour Glass Condensate (CGC) effective theory to describe the dynamics of a heavy-ion collision in the early stages after the collision ($\tau \lesssim 1$ fm). Many reviews of the CGC effective theory have been published, see for example \cite{Gelis:2010nm,Lappi:2010ek}.
The details of the evolution of a quark-gluon plasma (QGP) during this early stage are not well understood, but they are important because they provide the initial conditions for subsequent hydrodynamic evolution. 
The CGC approach is based on a separation of scales between valence partons with large nucleon momentum fraction, and  gluon fields with small nucleon momentum fraction. When the separation scale is fixed, the dynamics of the  gluon fields can be determined from the classical Yang-Mills (YM) equation with the source provided by the valence partons, by averaging over an ensemble of valence parton colour charge distributions. 

This paper is a companion paper to a previous work \cite{Carrington:2020ssh} in which we have explained the  strategy of our approach and some details of the calculational method. 
We use an expansion in the proper time, also called a `near field' expansion \cite{Fries:2005yc,Fukushima:2007yk,Fujii:2008km,Chen:2015wia,Fries:2017ina}. 
We work with infinitely Lorentz contracted sources, and the description of the system is classical.  
An advantage of the method however is that our results are analytic, and thus provide a potentially valuable alternative approach to the various numerical methods that are in use. 
In Ref. \cite{Carrington:2020ssh} we focused on the technical details of the calculation and showed only a few results, which were obtained for the  simple case of nuclei that are infinite in the transverse plane and uniform. In this paper we consider more physically realistic collisions where the nuclear area density function is not assumed constant. We present results from several different calculations and discuss their connection to experimental observables. 

In section \ref{sec-preliminaries} we define some notation and give a summary of the results of our previous paper. 
In section \ref{sec-analytic} we describe the structure of the energy-momentum tensor in Milne coordinates and, by exploiting the symmetries of the tensor, we give a fairly compact analytic expression for our result to order $\tau^2$. The order $\tau^4$ results are given in Appendix \ref{sec-order4}. 
We also formulate a calculation of the angular momentum of the glasma per unit rapidity, on a hypersurface of constant proper time. 
In section \ref{sec-numerical} we present some numerical results and discuss their significance in the context of  heavy-ion collisions at the Relativistic Heavy Ion Collider (RHIC) and the Large Hadron Collider (LHC). 
In section \ref{sec-notation} we define our notation for the geometry of the collision and the units we will use. 
In section \ref{sec-gradient} we discuss our method to implement non-constant nuclear area density functions, which is based on a gradient expansion of a two dimensional projection of a Woods-Saxon distribution.
We follow the method of \cite{Chen:2015wia} and discuss carefully the limitations of the expansion and how its convergence depends critically on what quantity is calculated. 
In sections \ref{ed-sec} and \ref{pressure-sec} we study the energy density and pressure of the glasma.
For a system in equilibrium the transverse and longitudinal pressures are equal to each other, and individually equal to a third of the energy density. 
A calculation of the pressure therefore gives information about how far the glasma is from equilibrium. 
We show that, for times small enough that the expansion converges, the transverse and longitudinal pressures move towards their equilibrium values.
We calculate a quantity that describes the asymmetry between the transverse and longitudinal pressures, and one that characterizes the difference between the two components of the transverse pressure. 
We show that the first is almost completely insensitive to the gradient expansion, but the second shows strong dependence. 
In section \ref{radial-flow-sec} we study the flow of the energy of the gluon field by calculating the radial projection of the Poynting vector. 
Our results in this section, and the previous two sections,  indicate that the expansion to order $\tau^6$ can be trusted to about $\tau=0.05$ fm.
In our previous paper \cite{Carrington:2020ssh} we have given a simple argument that this time is much greater than the lower bound at which we no longer
trust the classical description that is inherent in our approach.
In section \ref{sec-f-coefficients} we study the momentum anisotropy of the glasma by calculating the Fourier coefficients of the azimuthal distribution of the flow.
Anisotropy is an important characteristic because it is primarily sensitive to the properties of the system very early in its evolution. 
It is expected that the spatial anisotropy of the system that is produced at very early times (within the first few fm) will be encoded in the observed anisotropies of the final particle momentum distributions. Our results for the Fourier coefficients are larger than  expected, which does not agree with the common assumption that momentum anisotropy is mostly generated during the hydrodynamic evolution of the plasma. 
In section \ref{sec-ecc} we consider the spatial azimuthal asymmetry of the glasma by calculating the eccentricity. 
We look for correlations between spatial asymmetry and momentum anisotropy, which is characterized by the Fourier coeffients calculated in the previous section.
Since momentum anisotropy originates in the initial spatial asymmetries in the geometry of the system, such correlations 
provide information about the effect of the interactions, and the expansion of the system, during glasma formation. 
We find significant correlation between the eccentricity and the elliptic flow coefficient calculated in the previous section, which indicates that the spatial asymmetry of the initial energy density is converted into the anisotropy of the azimuthal distribution of the gluon momentum field, and this correlation is characteristic of hydro-like behaviour. 
In section \ref{sec-L2} we look at the angular momentum of the glasma, which develops in collisions with non-zero impact parameter, in the direction perpendicular to the reaction plane.
It has been proposed that because of spin-orbit coupling effects, the angular momentum of the glasma could lead to the
polarization of produced quarks and anti-quarks, which might be detected by measuring the polarization of the $\Lambda$ hyperon, or various vector mesons \cite{Liang:2004xn}. 
Measurements of this kind could be used to compare different hydrodynamic models and hadronization scenarios \cite{Liang:2019clf}.
Our results  are many orders of magnitude smaller than the initial angular momentum of two ions colliding with non-zero impact parameter. 
Our findings therefore disagree with the proposal that the glasma acquires a large fraction of the angular momentum of the participating valence quarks \cite{Liang:2019clf}, but is not contradicted by any experimental evidence, since attempts to measure the polarization of produced hadrons have found only very small effects  \cite{Adam:2018ivw,Acharya:2019ryw}.
In section \ref{sec-conclusions} we conclude with some discussion and comments on possible future directions of this research.

\section{Summary of previous work}
\label{sec-preliminaries}

\subsection{Energy-momentum tensor}
\label{sec-EMT}

In a previous paper \cite{Carrington:2020ssh} we calculated the energy-momentum tensor using the CGC effective theory, to sixth order in an expansion in the proper time. In this section, we review the main elements of that calculation. 

We consider a collision of two ions moving towards each other along the $z$-axis and colliding at $t=z=0$.
In the pre-collision region of space-time, the system is most naturally described using light-cone coordinates, but in the post-collision region Milne coordinates are more convenient. 
At the end of the calculation, when we look at physical quantities, we usually want to use Minkowski coordinates. 
The calculation is done most efficiently by using the three different coordinate systems at different stages, and transforming between them as needed. 
We will use greek letters for 4-indices, and latin letters for 2-indices denoting coordinates in the transverse plane. 
For example, the components of a position 4-vector in Minkowski coordinates $(t,z,x,y)$ are written $x^\mu$ with $\mu\in(0,1,2,3)$, 
and the transverse components are denoted $x^i$ with $i\in(2,3)$. 
In light-cone and Milne coordinates, the time and longitudinal coordinates are defined as functions of the Minkowski variables $(t,z)$, but the transverse coordinates are the same in all three bases. 
Our notation for light-cone and Milne variables is standard, and is reviewed in Appendix \ref{app-notation}.

The vector potential of the gluon field is described with the ansatz \cite{Kovner:1995ts} 
\bea
\label{ansatz}
&& A^+(x) = \Theta(x^+)\Theta(x^-) x^+ \alpha(\tau,\vec x_\perp)  \\
&& A^-(x) = -\Theta(x^+)\Theta(x^-) x^- \alpha(\tau,\vec x_\perp) \nonumber\\
&& A^i(x) = \Theta(x^+)\Theta(x^-) \alpha_\perp^i(\tau,\vec x_\perp)
+\Theta(-x^+)\Theta(x^-) \beta_1^i(x^-,\vec x_\perp)
+\Theta(x^+)\Theta(-x^-) \beta_2^i(x^+,\vec x_\perp)\,.\nonumber
\eea
The functions $\beta_1^i(x^-,\vec x_\perp)$ and $\beta_2^i(x^+,\vec x_\perp)$ represent the pre-collision potentials, and the functions $\alpha(\tau,\vec x_\perp)$ and $\alpha_\perp^i(\tau,\vec x_\perp)$ give the post-collision potentials. 
In the forward light-cone the vector potential satisfies the sourceless YM equation.
We find solutions valid for early post-collision times by expanding in the proper time $\tau$ \cite{Fries:2005yc,Fukushima:2007yk,Fujii:2008km,Chen:2015wia,Fries:2017ina}. Using these solutions we can write the post-collision field-strength tensor, and energy-momentum tensor, in terms of the initial potentials $\alpha(0, \vec x_\perp)$ and $\vec\alpha_\perp(0, \vec x_\perp)$.
The initial potentials are related to the physical properties of the ions and the geometry of the collision using boundary conditions that
connect the pre-collision and post-collision potentials. These conditions are found by integrating the YM equation across the light-cone \cite{Kovner:1995ts, Carrington:2020ssh} 
\bea
\label{cond1}
&& \alpha^{i}_\perp(0,\vec{x}_\perp) = \alpha^{i(0)}_\perp(\vec{x}_\perp) = \lim_{\text{w}\to 0}\left(\beta^i_1 (x^-,\vec{x}_\perp) + \beta^i_2
(x^+,\vec{x}_\perp)\right)\nonumber \\
&& \label{cond2}
\alpha(0,\vec{x}_\perp) = \alpha^{(0)}(\vec{x}_\perp) = -\frac{ig}{2}\lim_{\text{w}\to 0}\;[\beta^i_1 (x^-,\vec{x}_\perp),\beta^i_2
(x^+,\vec{x}_\perp)]\,\label{b-conds}
\eea
where the notation $\lim_{\text{w}\to 0}$ indicates that the width of the sources across the light-cone is taken to zero (the pre-collision potentials depend only on transverse co-ordinates in this limit). 
Using equations (\ref{ansatz}, \ref{b-conds}) the energy-momentum tensor can be written in terms of the pre-collision potentials $\vec \beta_1(x^-,\vec x_\perp)$ and $\vec \beta_2(x^+,\vec x_\perp)$ and their derivatives. 

The next step is to  use the YM equation to write the pre-collision potentials 
in terms of the colour charge distributions of the incoming ions.  
One then averages over a Gaussian distribution of colour charges within each nucleus.
The average of a product of colour charges can be written as a sum of terms that combine the averages of all possible pairs, which is called Wick's theorem. 
The average of a product of pre-collision potentials, which depend on the colour charges in a nontrivial way, is much more difficult to calculate, and the calculation becomes more and more complicated as the number of potentials increases \cite{Blaizot:2004wv, Fukushima:2007dy, Albacete:2018bbv, FillionGourdeau:2008ij, Lappi:2017skr}. 
We use the Glasma Graph approximation \cite{Lappi:2017skr}, which was also used in other near field expanded calculations, and is equivalent to the application of Wick's theorem to light-cone potentials directly. 
In our previous work \cite{Carrington:2020ssh} we showed that for the simple case of homogeneous ions that are infinite in the transverse plane, there is some evidence that the effect of this approximation is small. 
We stress however that the range of validity of the Glasma Graph approximation has not been carefully studied, and this is an open and important issue. 

In the context of our calculation, the use of the Glasma Graph approximation means that the energy-momentum tensor can be written in terms of the 2-point correlators of the pre-collision potentials. This correlator was originally calculated in Ref. \cite{JalilianMarian:1996xn} and generalized to include some effects of nuclear structure in Ref. \cite{Chen:2015wia}. 
We give below the result in our notation, see \cite{Carrington:2020ssh} for further details. 

The correlator of two potentials from different ions is assumed to be zero. 
The 2-point correlator for two potentials from the same ion is written in terms of the  colour charge surface density for that ion, which we denote $\mu_1(\vec x_\perp)$ and $\mu_2(\vec x_\perp)$. 
These functions are not determined by the CGC model, rather one assumes some form, constrained by experimental knowledge. 
We use a 2-dimensional projection of a Woods-Saxon distribution which is characterized by three parameters that correspond to the surface thickness, radius, and displacement of the centre of the ion relative to the beam axis (for details see Section \ref{sec-woods-saxon}).
The 2-point correlators for the two ions have the form
\bea
&&\delta^{ab} B_1^{ij}(\vec{x}_\perp,\vec y_\perp) \equiv 
 \lim_{{\rm w} \to 0}  \langle \beta_{1\,a}^i(x^-,\vec x_\perp) \beta_{1\,b}^j(y^-,\vec y_\perp)\rangle   \nn\\
&& \delta^{ab} B_2^{ij}(\vec{x}_\perp,\vec y_\perp) \equiv 
 \lim_{{\rm w} \to 0}  \langle \beta_{2\,a}^i(x^+,\vec x_\perp) \beta_{2\,b}^j(y^+,\vec y_\perp)\rangle   \,
\, \label{core5-20}
\eea
and using the index $n\in\{1,2\}$ to represent the two ions we have
\bea
&& B_n^{ij}(\vec{x}_\perp,\vec y_\perp) = g^2 \left(\frac{e^{g^4 N_c\;\Delta\tilde\gamma_n(\vec x_\perp,\vec y_\perp)}-1}
{g^4 N_c \Delta\tilde\gamma_n(\vec x_\perp,\vec y_\perp)}\right)  \,
\partial_x^i \partial_y^j \tilde\gamma_n(\vec x_\perp,\vec y_\perp) \,\label{B-res}
\eea
where
\bea
&& \Delta\tilde\gamma_n(\vec x_\perp,\vec y_\perp) = \tilde\gamma_n(\vec x_\perp,\vec y_\perp) -\frac{1}{2}\big( \tilde\gamma_n(\vec x_\perp,\vec x_\perp) + \tilde\gamma_n(\vec y_\perp,\vec y_\perp) \big) \label{Gamma-tilde-def} \\
&& \tilde \gamma_n(\vec x_\perp,\vec y_\perp) = \int d^2 z_\perp \, \mu_n(\vec z_\perp)\, G(\vec x_\perp-\vec z_\perp)\,G(\vec y_\perp-\vec z_\perp)\,
\label{gamma-tilde-def} \\
&& G(\vec x_\perp) = \frac{1}{2\pi} K_0(m |\vec x_\perp|)\,.
\label{Gf-def}
\eea
The function $K_0$ is a modified Bessel function of the second kind, and $m$ is an infrared regulator.
In the limit $m\to 0$ the Green's function $G(\vec x_\perp)$ goes like $\sim \ln(m |\vec x_\perp|)$. Since valence parton sources come from individual nucleons, the Green's function should go to zero when $|\vec x_\perp|$ approaches the confinement scale, and therefore we choose  $m \sim \Lambda_{\rm QCD}$. 

\subsection{Parton surface density}
\label{sec-woods-saxon}

To make further progress we must specify the form of the colour charge surface density of the nuclei. 
We will use a 2-dimensional projection of a Woods-Saxon potential of the form 
\bea
&& \mu(\vec x_\perp) \label{def-mu2-2} 
 =  \left(\frac{A}{207}\right)^{1/3}\frac{\bar\mu}{2a\ln(1+e^{R_A/a})} 
\int^\infty_{-\infty} dz \frac{1}{1 + \exp\big[(\sqrt{(\vec x_\perp)^2 + z^2} - R_A)/a\big]}\,.
\eea
The parameters $R_A$ and $a$ give the radius and skin thickness of a nucleus of mass number $A$, and their numerical values are discussed in section \ref{sec-gradient}. 
The integral in (\ref{def-mu2-2}) is normalized so that for a lead nucleus $\mu(\vec 0)=\bar\mu$, which is sometimes called the McLerran-Venugopalan (MV) scale. This parameter is related to the saturation scale $Q_s$, but its exact value cannot be determined within the CGC approach (for a discussion see \cite{Iancu:2003xm}). We use $\bar\mu = Q_s^2/g^4$, and some motivation for this choice can be found in \cite{Carrington:2020ssh}. 
Due to the ambiguity associated with the value of the MV scale, our numerical results for quantities like the energy density and pressure should be regarded as order of magnitude estimates. Quantities that depend on ratios of different elements of the energy momentum tensor, like Fourier coefficients of the azimuthal flow, will have much weaker dependence on the MV scale. 

We obtain an analytic result for the energy-momentum tensor by substituting equation (\ref{def-mu2-2}) into equation (\ref{gamma-tilde-def}) and performing a gradient expansion, using the method developed in Ref. \cite{Chen:2015wia}. 
The coordinates $\vec x_\perp$ and $\vec y_\perp$ are rewritten in terms of  relative and average coordinates.
To consider collisions with non-zero impact parameter we expand the distribution $\mu_1(\vec z_\perp)$ around the average coordinate $(\vec x_\perp+\vec y_\perp)/2 = \vec R-\vec b/2$, and $\mu_2(\vec z_\perp)$ around $(\vec x_\perp+\vec y_\perp)/2 = \vec R+\vec b/2$. 
We will keep terms up to second order in gradients of the distribution. The parameter that we assume to be small is
\bea
\delta = \frac{|\nabla^i\mu(\vec R \pm \frac{\vec b}{2})|}{m \mu(\vec R \pm \frac{\vec b}{2})} \,
\label{delta-def}
\eea
where the gradient operator indicates differentiation with respect to the argument of the function. The region of validity of this expansion is discussed in section \ref{sec-gradient}. 

We remind the reader that for a realistic nucleus, which is made up of individual partons, 
the transverse charge distribution is not a very smooth function. 
It is possible that the transverse charge distribution of a real nucleus could be sufficiently irregular that a Woods-Saxon distribution is not a good representation. 

\subsection{2-particle correlators}
\label{sec-corr}

In this subsection we drop the subscript that indicates which ion is being considered, and we set $\vec b=0$. 
Performing the derivative expansion and keeping terms up to second order in gradients of $\mu$,  equation (\ref{gamma-tilde-def})  becomes
\bea
 \tilde\gamma(\vec x_\perp,\vec y_\perp) &=& \frac{\mu(\vec R)r}{4\pi m}K_1(mr)+\frac{1}{2}\nabla^i \nabla^j \mu(\vec R) \left(\delta^{ij}\frac{r^2}{24\pi m^2}K_2(mr)+\frac{r^i r^j}{r^2}\frac{r^3}{48\pi m}K_1(mr)\right)\,.
 \nonumber\\
 \label{gamma-def-2}
\eea
We can rewrite equation (\ref{gamma-def-2}) in the form 
\bea
\tilde \gamma(\vec x_\perp,\vec y_\perp) &=& \mu(\vec R) \int \frac{d^2k}{(2\pi)^2} \frac{e^{i\vec r\cdot \vec k} }{\left(k^2+m^2\right)^2} 
+ \frac{m^2}{2} \nabla ^2\mu(\vec R)  \int \frac{d^2k}{(2\pi)^2} \frac{e^{i\vec r\cdot \vec k} }{\left(k^2+m^2\right)^4}\,
\label{gamma-def-3}
\eea
where we have made the replacement $\hat r^i \hat r^j \to \delta^{ij}/2$ because in the limit $\vec r\to 0$ we know $\tilde\gamma$ must be independent of the direction of the vector $\hat r$. We note we are able to make this replacement before performing any derivatives with respect to $\vec x_\perp$ and $\vec y_\perp$,  since $\lim_{r\to 0}\partial_x^i \dots \partial_y^j \dots \hat r^k \hat r^l = 0$, where the dots indicate any number of derivatives. 

The correlator $B^{ij}(\vec x_\perp, \vec y_\perp)$ and its derivatives  have ultra-violet divergences that must be regulated.
We use a modified version of the method proposed in Ref. \cite{Fujii:2008km}. 
To illustrate how we make use of equation (\ref{gamma-def-3}) we consider, for example, the calculation of 
\bea
\partial_x^i\partial_y^j \tilde\gamma(\vec x_\perp,\vec y_\perp)&=& \mu(\vec R) \int \frac{d^2k}{(2\pi)^2} \frac{k^i k^j  e^{i\vec r\cdot \vec k} }{\left(k^2+m^2\right)^2} 
+ \frac{m^2}{2} \nabla ^2\mu(\vec R)  \int \frac{d^2k}{(2\pi)^2} \frac{k^i k^j  e^{i\vec r\cdot \vec k} }{\left(k^2+m^2\right)^4}
\label{bxx-example}
\eea
which appears in the expression for $B^{ij}$ in equation (\ref{B-res}). 
The integration over angular variables gives $k^i k^j \to \delta^{ij} k^2/2$. 
The second term in (\ref{bxx-example}) is finite, but the first term is logarithmically divergent and we regulate it using an ultraviolet momentum cutoff $\Lambda$. 
This cutoff will be set to the saturation scale, see our previous paper \cite{Carrington:2020ssh} for some discussion of this point. In section \ref{sec-martinez} we test the dependence of some of our results on the saturation scale, and show that they are not very sensitive to its numerical value.

Now we consider the contribution from the factor in round brackets in equation (\ref{B-res}). Expanding this factor we have
\bea
\frac{e^{g^4 N_c\;\Delta\tilde\gamma_n(\vec x_\perp,\vec y_\perp)}-1}
{g^4 N_c \Delta\tilde\gamma_n(\vec x_\perp,\vec y_\perp)} = 1+\frac{1}{2} g^4 N_c \Delta\tilde\gamma(\vec x_\perp,\vec y_\perp) 
+ \frac{1}{6} \left(g^4 N_c \Delta\tilde\gamma(\vec x_\perp,\vec y_\perp)\right)^2 + \dots
\label{coef-exp}
\eea 
When we calculate derivatives of the correlator $B^{ij}(\vec x_\perp,\vec y_\perp)$, the derivatives operate on all terms in the expansion in equation (\ref{coef-exp}). At sixth order in the $\tau$ expansion the energy-momentum tensor includes terms with six derivatives acting on the correlator in equation (\ref{B-res}). 
Naively it would seem that we need to expand the exponent in equation (\ref{coef-exp}) to seventh order, since each of the six derivative operators
will have a piece proportional to $\partial/\partial r^i$ which could act separately on each of the six factors
in the term $(\Delta\tilde\gamma(\vec x_\perp,\vec y_\perp))^6$. For example, if we differentiate six times with respect to $r^1$ we obtain an expression of the form
\bea
\lim_{\vec r\to 0} \left(\frac{\partial}{\partial r^1}\right)^6 (\Delta\tilde\gamma(\vec x_\perp,\vec y_\perp))^6 = \lim_{\vec r\to 0} \left(\frac{\partial}{\partial r^1}\tilde\gamma(\vec x_\perp,\vec y_\perp)\right)^6 + \dots
\eea
where the dots represent additional terms that give zero when $\vec r$ is taken to zero. 
However, it is easy to see from equation (\ref{gamma-def-3}) that if we differentiate $\tilde\gamma$ an odd number of times with respect to $r^1$ or $r^2$, the integration over momentum variables gives zero. 
This means terms with more than three factors of  $\Delta\tilde\gamma$, which are operated on with a maximum of six derivatives with respect to components of $\vec r$, can be set to zero. Equivalently, we have to expand the exponential only to fourth order. 

All correlators and their derivatives can be obtained using the method described above. We give one example
\bea
&& \lim_{r\to 0}B^{ij}(\vec x_\perp,\vec y_\perp) 
=\delta^{ij}g^2\frac{\mu(\vec R)}{8\pi}\left[\ln\left(\frac{\Lambda^2}{m^2}+1\right)-\frac{\Lambda^2}{\Lambda^2+m^2}\right] \nn\\
&& + \frac{g^2}{16\pi(\Lambda^2+m^2)} \left[\delta^{ij} \nabla^2\mu(\vec R)\frac{\Lambda^4}{6m^2(\Lambda^2+m^2)}\left(1+\frac{2m^2}{\Lambda^2+m^2}\right)
+ \nabla^i\nabla^j \mu(\vec R)\frac{\Lambda^2}{m^2}\right]\,.
\label{g-corr-res-basic}
\eea

\section{Analytic Results}
\label{sec-analytic}

\subsection{Structure of the energy-momentum tensor}
\label{sec-structure}
In this section we present our analytic result for the energy-momentum tensor, to order $\tau^6$. 
For simplicity of notation we give our results for the tensor in Milne coordinates, where there is no dependence on rapidity. 
All elements in the energy-momentum tensor have either even or odd powers of $\tau$. 
We summarize this information in the symbolic equation
\bea
{\cal O}\big(T_{\rm milne}\big) = \left(
\begin{array}{cccc}
 ~~~(0,2,4,6) ~~~& ~~~ (1,3,5)~~~ & ~~~(1,3,5)~~~ & ~~~(1,3,5) \\
 (1,3,5) & (-2,0,2,4) & (0,2,4) & (0,2,4) \\
 (1,3,5) & (0,2,4) & (0,2,4,6) & (2,4,6) \\
 (1,3,5) & (0,2,4) & (2,4,6) & (0,2,4,6) \\
\end{array}
\right).
\eea
The top left element of the tensor shows that the element $T^{00}$ has contributions of order $\tau^0$, $\tau^2$, $\tau^4$ and $\tau^6$. 
To give one other example, the entry in the top right corner shows that the element $T^{03}$ has contributions of order $\tau$, $\tau^3$ and $\tau^5$. 

We do not need to give all components of the energy-momentum tensor because of its symmetry properties. 
We will give our results for the six elements in equation (\ref{T-symall}) that are written in  boldface.  
All of the other elements in the tensor can be generated from these using symmetries, as explained below. 
We emphasize that we have calculated all elements of the energy-momentum tensor, and the symmetries summarized in equation (\ref{T-symall}) have been verified, and not assumed. 

All elements below the diagonal can be related to elements above the diagonal using the fact that the tensor is symmetric. In equation (\ref{T-symall}) we write, for example, $T^{10} = T^{01}$. 
There are also pairs of elements where one can be obtained from the other using the transformation $\nabla_x \leftrightarrow \nabla_y$ where we have defined
$\nabla_x\equiv \partial/\partial R^x$ 
and $\nabla_y\equiv \partial/\partial R^y$. We will also use $\nabla^2 \equiv \nabla_x^2+\nabla_y^2$.
For example, we write $T^{13} = {\cal F}T^{12}$ to indicate that the element $T^{13}$ can be obtained from $T^{12}$ by interchanging the derivative operators $\nabla_x$ and $\nabla_y$.
Finally, since the energy-momentum tensor is traceless it satisfies $g_{\mu\nu}T^{\mu\nu}=0$, which means that we do not have to give all elements on the diagonal. We replace the element $T^{11}$ with the symbol ${\it Tr}$ to indicate that this matrix element can be constructed from the other diagonal elements of the tensor. Combining this information we write
\bea
T_{\rm milne} \longrightarrow \left(
\begin{array}{cccc}
~~ {\bf T^{00}}~~ & ~~ {\bf T^{01}} ~~& ~~ {\bf T^{02}}~~ &~~ {\cal F}T^{02} \\
 T^{01} & {\it Tr} & {\bf T^{12}} & {\cal F}T^{12} \\
 T^{02} & T^{12} & {\bf T^{22}} & {\bf T^{23}} \\
 T^{03} & T^{13} & T^{23} & {\cal F}T^{22} \\
\end{array}
\right) \,.\label{T-symall}
\eea
In summary, equation (\ref{T-symall}) tells us that we need to give only the six elements of the energy-momentum tensor that are written in boldface.

\subsection{Coefficients of the energy-momentum tensor}
\label{sec-order2}

We give generic equations for these elements organized by powers of $\tau$ and numbers of derivatives with respect to each of the transverse coordinates.
Our result for the energy-momentum tensor has the form
\bea
&& T^{00} = {\cal E}_0^{00}+ (1+{\cal F}){\cal E}_0^{02} 
+ \tau^2\big(\frac{1}{6}\nabla^2 {\cal E}_0^{00} + {\cal E}_2^{00}+ (1+{\cal F}){\cal E}_2^{20}\big) \nn\\
&&~~ +\tau^4\big(\frac{1}{10}\nabla^2 {\cal E}_2^{00} + {\cal E}_4^{00} + (1+{\cal F}){\cal E}_4^{02} \big)
+\tau^6\big({\cal E}_6^{00} + (1+{\cal F}) {\cal E}_6^{02}  \big) \nn\\
&& T^{01} = -\frac{1}{8}(1+{\cal F})\left(2\tau \nabla_x \beta^{10}_0 
+ \frac{4}{3}\tau^3 \nabla_x \beta^{10}_2 +\tau^5 \nabla_x \beta^{10}_4 \right)  \nn\\
&& T^{02} = -\frac{1}{2}\tau \nabla_x {\cal E}_0^{00} -\frac{1}{2}\tau^3 \nabla_x {\cal E}_2^{00} -\frac{1}{2}\tau^5 \nabla_x {\cal E}_4^{00} \nn\\
&& T^{12} = \beta^{10}_0 + \tau^2 \beta^{10}_2 + \tau^4 \beta^{10}_4 \nn\\
&&T^{22} = {\cal E}_0^{00}+ (1+{\cal F}){\cal E}_0^{02}
 + \tau^2 \left(2{\cal E}_2^{00} + \delta_2^{20} + \delta_2^{02} \right)
+\tau^4(3{\cal E}_4^{00}+\delta_4^{20}+\delta_4^{02}) 
+\tau^6(4{\cal E}_6^{00}+\delta_6^{20}+\delta_6^{02}) \nn\\
&& T^{23} = \tau^2 \gamma_2^{11} + \tau^4 \gamma_4^{11} + + \tau^6 \gamma_6^{11} \,. \label{TTgeneric7}
\eea   
For each of the greek letter variables in equation (\ref{TTgeneric7}) the subscript gives power of $\tau$ that multiplies the variable when it is not acted on by any external derivatives, and the superscripts give the number of internal derivatives with respect to $R_x$ and $R_y$. For example, the second term in $T^{12}$ contains the variable $\beta^{10}_2$ which has coefficient $\tau^2$ and is defined (see equation (\ref{results_order2})) as a sum of terms each of which has one derivative with respect to $R_x$. 

We use ${\cal X}$ to denote any of the greek letters $\{{\cal E}, \beta, \gamma, \delta, \xi \}$.
We give below our results for  ${\cal X}_n^{lm}$ with $0\le n \le 2$.
In Appendix \ref{sec-order4} we give our results for   ${\cal X}_n^{lm}$ with $n=4$.
Each symbol is either even or odd under the transformation $\mu_1 \leftrightarrow \mu_2$, and therefore we only need to give half of the terms, and the sign of the symmetry. 
We set the ultraviolet cutoff $\Lambda$ equal to the saturation scale $Q_s$ and introduce the notation $L\equiv\ln(Q_s/m)$. 
To save space we give only the contributions that are leading-order in an expansion in the infrared mass $m$, and this power counting is done with the assumption $(\nabla_x)^{n_i}(\nabla_y)^{n_j} \mu_1(\vec R)/\mu_1(\vec R) \sim m^{n_i+n_j}$ (and similarly for $\mu_2(\vec R)$). 
We will also factor the constant $\bar\mu\equiv Q_s^2/g^4$ out of the source density functions and define
$\hat\mu_1(\vec R) \equiv \mu_1(\vec R)/\bar\mu $, 
and similarly for ion 2.
To save space we will use the shorthand notation $\hat\mu_1^{10} \equiv \nabla_x \hat\mu_1(\vec R)$, $\hat\mu_1^{11} \equiv \nabla_x \nabla_y \hat\mu_1(\vec R)$, etc.
We emphasize that in all of our numerical calculations there is no expansion in $m$ and we include all contributions from the gradient expansion up to second order.  

Using the notation defined above, the coefficients of the energy-momentum tensor at order $\tau^0$  have the simple form
\bea
{\cal E}_0^{00} &=& \frac{3 \hat\mu _1 \hat\mu _2 Q_s^4}{8 \pi ^2 g^2} (2 L-1)^2 + (\hat\mu_1 \leftrightarrow \hat\mu_2) \nn\\
\beta_{0}^{10} &=& 
\frac{3 \hat\mu _1 Q_s^4 \hat\mu _2{}^{10}}{8 \pi ^2 g^2}(2 L-1)^2 - (\hat\mu_1 \leftrightarrow \hat\mu_2) \nn\\
{\cal E}_0^{02} &=& \frac{\hat\mu _1 Q_s^4 \hat\mu _2{}^{02}}{4 \pi ^2 g^2 m^2}  
(2L-1) + (\hat\mu_1 \leftrightarrow \hat\mu_2)\,.
\label{results_order0}
\eea
At order $\tau^2$ we have
\bea
\beta_2^{10} &=& 
\frac{27 \hat\mu _1 Q_s^6}{512 \pi ^3 g^2}
\bigg(
(2 L-1) \left(11 \hat\mu _2 (1-2 L)^2 \hat\mu _1{}^{10}+\left(11 \hat\mu _1 (1-2 L)^2-8 \pi \right) \hat\mu _2{}^{10}\right)
\bigg) - (\hat\mu_1 \leftrightarrow \hat\mu_2) \nn \\
\gamma_2^{11} &=& 
\frac{9 \hat\mu _1 Q_s^6}{512 \pi ^3 g^2 m^2}
\bigg(
(1-2 L)^2 \left(8 \hat\mu _2 \hat\mu _1{}^{11}+7 \hat\mu _1 \hat\mu _2{}^{11}\right)
\bigg) + (\hat\mu_1 \leftrightarrow \hat\mu_2) \nn \\
\delta_2^{02}  &=& -\frac{\hat\mu _1 Q_s^6}{1024 \pi ^3 g^2 m^2}
\bigg(
750 \hat\mu _2 (2 L-1)^2 \hat\mu _1{}^{02}+\left(447 \hat\mu _1 (2 L-1)^2+128 \pi \right) \hat\mu _2{}^{02}
\bigg) + (\hat\mu_1 \leftrightarrow \hat\mu_2) \nn\\
\delta_2^{20}  &=& 
-\frac{\hat\mu _1 Q_s^6}{1024 \pi ^3 g^2 m^2}
\bigg(
606 \hat\mu _2 (1-2 L)^2 \hat\mu _1{}^{20}+\left(321 \hat\mu _1 (1-2 L)^2+128 \pi \right) \hat\mu _2{}^{20}
\bigg) + (\hat\mu_1 \leftrightarrow \hat\mu_2) \nn\\
{\cal E}_2^{00}  &=& 
-\frac{3 \hat\mu _1 \hat\mu _2 (2 L-1) Q_s^6}{16 \pi ^3 g^2}
\bigg(
3 \hat\mu _2 (2L-1)^2+\pi
\bigg) + (\hat\mu_1 \leftrightarrow \hat\mu_2) \nn\\
{\cal E}_2^{02}  &=& 
-\frac{\hat\mu _1 Q_s^6}{1024 \pi ^3 g^2 m^2}
\bigg(
339 \hat\mu _2 (2L-1)^2 \hat\mu _1{}^{02}+64 \left(3 \hat\mu _1 (2 L-1)^2+\pi \right) \hat\mu _2{}^{02}
\bigg) + (\hat\mu_1\leftrightarrow \hat\mu_2) \,.\nn\\
\label{results_order2}
\eea

\subsection{Angular momentum}
\label{sec-L}

In this section we derive an expression for the angular momentum of the glasma per unit rapidity. 
Our method is similar to that of Ref. \cite{Fries:2017ina}. 
We define the tensor 
\bea
M^{\mu\nu\lambda}=T^{\mu\nu}R^\lambda - T^{\mu\lambda}R^\nu \,
\label{M-def}
\eea
where $R^\mu$ denotes a component of the position vector.
The energy momentum tensor is divergenceless, and therefore the tensor in equation (\ref{M-def}) satisfies the tensor equation $\nabla_\mu M^{\mu\nu\lambda} = 0$. 
Using Stokes' theorem one obtains a set of six conserved quantities
\bea
J^{\nu\lambda} = \int_\Sigma d^3 y \sqrt{|\gamma|} n_\mu M^{\mu\nu\lambda}\,
\label{J1}
\eea
where $n^\mu$ is a unit vector perpendicular to the hypersurface $\Sigma$, $\gamma$ is the induced metric on this hypersurface, and $d^3y$ is the corresponding volume element.
The angular momentum is obtained from the Pauli-Lubanski vector 
\bea
L_\mu = -\frac{1}{2}\epsilon_{\mu\alpha\beta\gamma}J^{\alpha\beta}u^\gamma
\label{L-def}
\eea
where $u^\gamma$ is the vector that denotes the rest frame of the system. 
One can easily verify that equations (\ref{M-def}, \ref{J1}, \ref{L-def}) reduce to the usual definition of angular momentum in Minkowski space. 
We denote indices for spatial variables in Minkowski space with underscored latin letters, for example $\underline{i}\in(1,2,3)$ and $x^{\underline{i}}$ is a component of the vector $(x,y,z)$. 
We use $n^\mu = (1,0,0,0)$ so that $\Sigma$ is a hypersurface of constant $t$,
 and with $u^\gamma = (1,0,0,0)$ equation (\ref{L-def}) becomes
\bea
L^{\underline{i}}_{\rm mink} = -\frac{1}{2}\epsilon^{\underline{i}\underline{j}\underline{k}}J^{\underline{j}\underline{k}} = -\frac{1}{2} \epsilon^{\underline{i}\underline{j}\underline{k}} \int d^3 \vec x \,(T^{0\underline{j}}x^{\underline{k}} - T^{0\underline{k}}x^{\underline{j}}) = \epsilon^{\underline{i}\underline{j}\underline{k}} \int d^3 \vec x \, x^{\underline{j}} P^{\underline{k}}\,
\label{L-mink}
\eea
where $d^3 \vec x$ represents the spatial volume element in Minkowski space, and we have written the Poynting vector $P^i\equiv T^{0i}$.

We work in Milne coordinates and use $n^\mu = (1,0,0,0)$ so that 
\bea
J^{\nu\lambda} = \tau \int d\eta\,d^2 \vec R \, M^{0\nu\lambda}\, \label{J-now}
\eea
is defined on a hypersurface of constant $\tau$. 
Using $u^\gamma = (1,0,0,0)$  gives
\bea
L_\mu = \frac{1}{2}\tau \, \epsilon_{0\mu\alpha\beta} \int d\eta \int d^2\vec R \, (T^{0\alpha}R^\beta-T^{0\beta}R^\alpha)\,.\label{L-def-2}
\eea
In our calculation angular momentum is not conserved, because the currents on the light-cone act as sources, and therefore we consider instead the angular momentum per unit rapidity. 
From equation (\ref{L-def-2}) we obtain\footnote{We comment that our result is different from that of Ref. \cite{Fries:2017ina} where a slightly inconsistent procedure is used. 
In that paper the authors define angular momentum in Minkowski space, on a surface of constant time, and then enforce separately that the integral should be calculated with $\tau$ held fixed.}
\bea
\frac{dL_\mu}{d\eta} = \frac{1}{2}\tau \, \epsilon_{0\mu\alpha\beta}  \int d^2\vec R \, (T^{0\alpha}R^\beta-T^{0\beta}R^\alpha)\,.\label{L-def-3}
\eea
We note that although the right side of  equation (\ref{L-def-3}) is independent of rapidity, our calculation is only meaningful close to mid-rapidity where boost invariance is a good approximation.

The integral over the transverse plane in equation (\ref{L-def-3}) can be simplified using symmetry considerations. 
The source charge distributions that we use (see Section \ref{sec-woods-saxon}) are even under the transformation $R^y \to -R^y$. 
We will consider symmetric displacements of the ions relative to the collision axis ($\vec b_1 = -\vec b_2$) so that the transformation $R^x \to -R^x$ interchanges the distributions for the first and second ions. 
Using these symmetries one can show that each component of the energy momentum tensor in Milne coordinates is either even or odd under the $R^x$ and $R^y$ parity transformations. We summarize these symmetries in Table \ref{symmetry-table}.
\begin{table}[H]
\centering 
\begin{tabular}{|c | c | c|} 
\hline                   
~ ~~~~~~ ~ & ~~~~ $R_x \to -R_x$ ~~~~ &~~~~ $R_y \to -R_y$ ~~~~   \\ [0.5ex]
\hline
$T^{00}$ & even & even   \\
$T^{01}$ & odd & even   \\
$T^{02}$ & odd & even   \\
$T^{03}$ & even & odd   \\
$T^{11}$ & even & even   \\
$T^{12}$ & even & even   \\
$T^{13}$ & odd & odd   \\
$T^{22}$ & even & even   \\
$T^{23}$ & odd & odd   \\
$T^{33}$ & even & even   \\
\hline
\end{tabular}
\caption{Symmetries of the components of the energy-momentum tensor under the transformations $R_x\to -R_x$ and $R_y\to -R_y$. }
\label{symmetry-table}
\end{table}
Using the symmetry relation in Table \ref{symmetry-table} it is easy to see that the only non-zero component of the angular momentum per unit rapidity is
\bea
\frac{dL^y}{d\eta} = -\tau^2 \int d^2\vec R \, R^x T^{01}\,.
\label{L-def-4}
\eea

\section{Numerical Results}
\label{sec-numerical}

\subsection{Notation and units}
\label{sec-notation}

We remind the reader of the geometry of the collision we are considering. The two ions approach each other along the $z$-axis and collide at the origin, at time $t=0$.
Post collision, the first ion moves outward along the positive $z$-axis, and the second ion moves along the negative $z$-axis.
We will consider collisions with non-zero impact parameter, which we denote $b$. The displacement vector for the first ion is $\vec b_1 = (b/2,0)$ and for the second ion we use $\vec b_2 = (-b/2,0)$. 
Energy and pressure are given in GeV  and lengths in fm (we use natural units $c=\hbar=1$).
We define a dimensionless time variable $\tilde \tau = \tau Q_s$. 
We use  $N_c=3$, $m=0.2$ GeV, $Q_s=2$ GeV and $g=1$, unless stated otherwise. 
We consider lead-lead collisions, which corresponds to mass numbers $A_1=A_2=207$, except for a few situations where we will explicitly specify different mass numbers. 
We will show below that our results to order $\tau^6$ are valid to $\tau\approx 0.05$ fm, which corresponds to $\tilde\tau\approx 0.51$.

\subsection{Physical observables and the gradient expansion}
\label{sec-gradient}

\begin{figure}[H]
\centering
\includegraphics[scale=0.25]{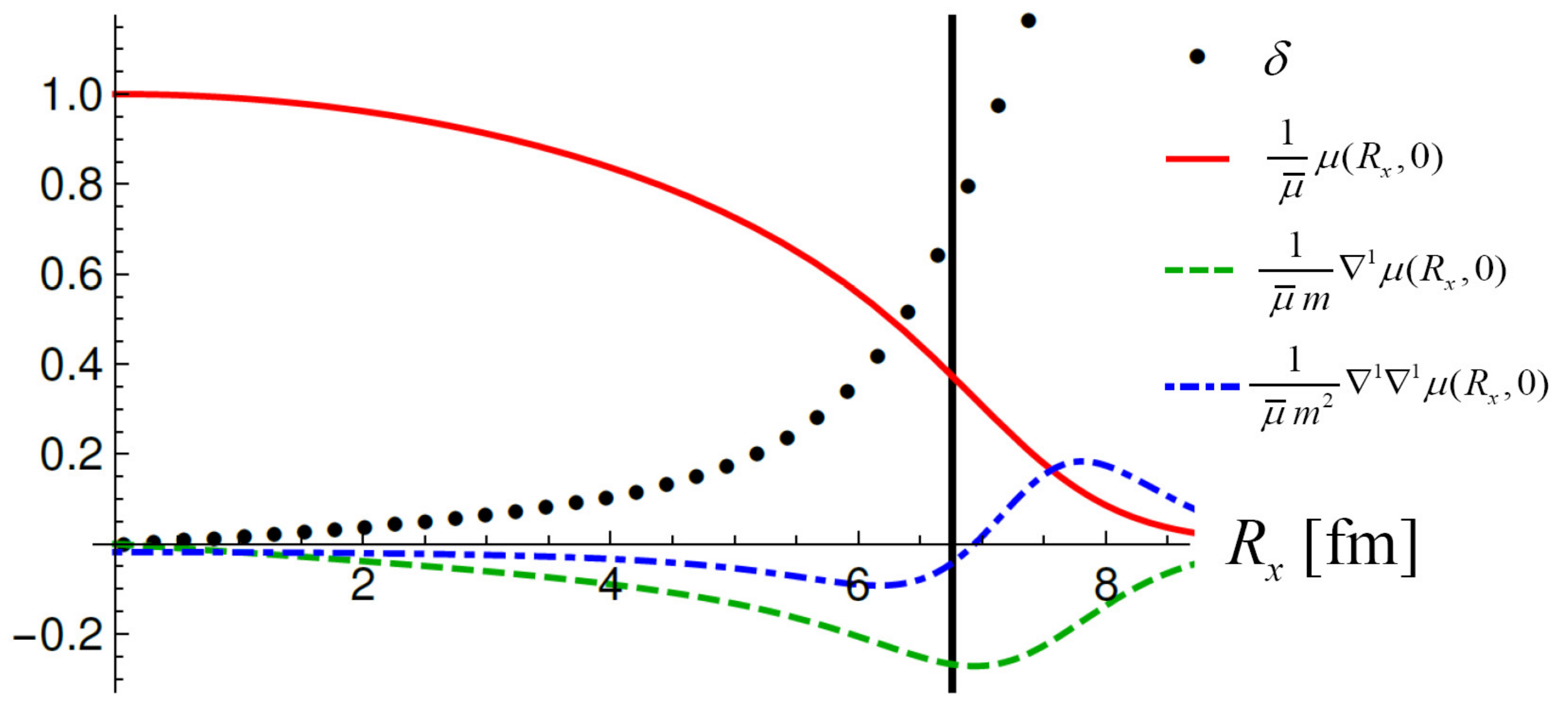}
\caption{The red (solid), green (dashed) and blue (dot-dashed) curves show the density function and its first and second derivatives. The quantity $\delta$ in equation (\ref{delta-def}) is shown by the black dots. For illustration the figure shows a vertical line that indicates the value of $R_x$ for which $\delta=0.75$. 
\label{MUd-plot}}
\end{figure}

The form of the charge density function $\mu(\vec x_\perp)$ that we use is discussed in section \ref{sec-woods-saxon}. 
We use $r_0=1.25$ fm and $a=0.5$ fm so that the radius of a nucleus with $A=207$ is $R_A=r_0 A^{1/3}= 7.4$ fm. 
As explained in section \ref{sec-woods-saxon}, we allow for non-homogeneous nuclear density functions by performing a gradient expansion around the coordinate that gives the position of the center of each nucleus in the transverse plane. 
In figure \ref{MUd-plot} we show the density function with $\vec x_\perp = (R_x,0)$, its first and second derivatives with respect to $R_x$, and the quantity $\delta$ in equation (\ref{delta-def}) which  must be small for the gradient expansion to converge. 
The condition $\delta < 0.75$ is satisfied in the region to the left of the vertical line in the figure.
Figure \ref{MUd-plot} shows clearly that the derivatives of the density function are appreciable only in a very small region at the edges of the nucleus. 
This means that if we calculate a quantity for which the dominant part of the integrand is not close to the edges of the nuclei, the gradient expansion will converge well, but the contributions of the derivative terms will likely be so small that they are negligible. 
On the other hand, if we calculate a quantity for which the region of the transverse plane close to the edges of the nuclei is important, the contribution from the derivative terms can be large, but the convergence of the gradient expansion must be studied carefully. 

In section \ref{ed-sec} we look at the energy density of the glasma.
We restrict to the region of the transverse plane for which $-5 \text{~fm} <|\vec R|<  5 \text{~fm}$. 
The corresponding condition on the expansion parameter is $\delta\lesssim 0.2$, and the gradient expansion converges well. 
Within this region, the inhomogeneity of the energy density in the transverse plane is almost entirely due to the asymmetry created by a non-zero impact parameter, which produces an almond shaped region of overlap. The gradients of the individual charge distributions are small and mostly irrelevant. 
In section \ref{pressure-sec} we calculate the pressure of the glasma.
To illustrate the issues associated with the gradient expansion we look at two different quantities, one of which describes the asymmetry between the transverse and longitudinal pressures, and the other  characterizes the difference between the two components of the transverse pressure. 
We show that the former is almost completely insensitive to the gradient expansion, whereas the latter depends strongly on the behaviour of the nuclear density function $\mu(\vec x_\perp)$ close to the nuclear radii.
In section \ref{radial-flow-sec} we look at various projections of the Poynting vector which describes the flow of the energy of the gluon field. 
The leading-order contribution to the Poynting vector comes solely from the first order term in the gradient expansion, which means that we can restrict to the region where the expansion parameter $\delta$ is small, and still see clearly the contribution of the gradient terms. 
In section \ref{sec-f-coefficients} we study the momentum anisotropy of the glasma by calculating the Fourier coefficients of the azimuthal distribution of the flow, and in  section \ref{sec-ecc} we look at the spatial azimuthal asymmetry of the glasma by calculating the eccentricity. 
These calculations involve integration over the transverse plane, and are therefore potentially sensitive to the gradient expansion. 
One must show that results are largely insensitive to the choice of the integration limits, and we will find that this condition restricts us to the consideration of small impact parameters. 
This is because when the centers of the two ions are separated, the inner edge of the first/second ion, where the density changes rapidly, will be closer to the center of the second/first ion, where integrand can be large. 
In section \ref{sec-L2} we look at the angular momentum of the glasma. 
In this case the gradient expansion severely limits the accuracy of the calculation, but we are able to see that the angular momentum carried by the glasma is many orders of magnitude smaller than the total angular momentum of the participant nucleons of the colliding nuclei \cite{Gao:2007bc, Becattini:2007sr}. 

\subsection{Energy density}
\label{ed-sec}

\begin{figure}[H]
\centering
\includegraphics[scale=1.6]{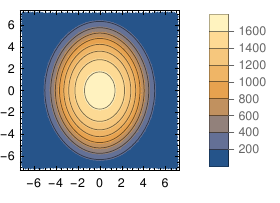} \quad
\includegraphics[scale=1.6]{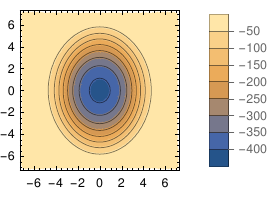} \quad
\caption{Energy densities in the transverse plane for case $B$. The left panel shows the energy density at $\tilde\tau=0$ and the right panel shows the difference between the energy densitiy at $\tilde\tau=0.42$ and the initial energy density. The units are GeV/fm$^3$ and the axes show $R_x$ and $R_y$ in fm. \label{fig-alm} }
\end{figure}

We look at the initial energy density ${\cal E} = T_{\rm mink}^{00}$ at mid-spatial-rapidity ($\eta=0$) for four different configurations of the colliding ions which are defined in Table \ref{config-table}. 
The last row of the table shows the maximum initial energy density. 
\begin{table}[H]
\centering 
\begin{tabular}{|c | c | c|c|c|c|c|c|c|c|c|} 
\hline                        
~ ~~~~~~ ~ & ~~~~ $A$ ~~~~ &~~~~ $B$ ~~~~ &~ ~~ ~ $C$ ~~~~ &~ ~~~ $D$ ~~~~   \\ [0.5ex]   
\hline
$A_1$ & 207 & 207 & 207 & 207  \\
$A_2$ & 207 & 207 & 40  & 40   \\
$b_1/2$ & 0 & 3 &  3 & 0  \\
$b_2/2$ & 0 &-3 & -3 & 0     \\
\hline
 ${\cal E}^{\rm max}_0$  GeV/fm$^3$ &  2080 & 1715 & 722 & 1202 \\  
\hline
\end{tabular}
\caption{Configurations of colliding ions}
\label{config-table}
\end{table}
In figure \ref{fig-alm} we show, for case $B$, the initial energy density, and the difference between the energy density at $\tilde\tau=0.42$ and the initial energy density. The energy density drops fastest at the centre and more slowly at the edges of the almond shaped interaction region. 

\subsection{Pressures}
\label{pressure-sec}

\subsubsection{Transverse and longitudinal pressures}
\label{sec-martinez}

We define the normalized longitudinal and transverse pressures as
\bea
\frac{p_L}{{\cal E}} = \frac{T_{\rm mink}^{11}}{T_{\rm mink}^{00}} \text{~~ and ~~} \frac{p_T}{{\cal E}} = \frac{1}{2}\frac{(T_{\rm mink}^{22} + T_{\rm mink}^{33})}{T_{\rm mink}^{00}}\,.
\eea
For a system in equilibrium  $p_T/{\cal E}=p_L/{\cal E}=1/3$. 
The glasma energy-momentum tensor at $\tau=0^+$ has the diagonal form (with both indices raised)
\bea
T_{\rm mink}^{\rm initial} = \left(
\begin{array}{cccc}
{\cal E}_0  & 0 & 0 & 0 \\
0 & - {\cal E}_0  & 0 & 0  \\
0 & 0 & {\cal E}_0  & 0  \\
0 & 0 & 0 & {\cal E}_0  
\end{array}
\right)\,.\label{diag}
\eea
We remind the reader that in our notation the components of a position 4-vector in Minkowski coordinates are $(t,z,x,y)$. Equation (\ref{diag}) shows that the initial longitudinal pressure is large and negative. 
The initial system is therefore not only far from equilibrium, but also far from the regime where a quasi-particle picture would be valid. 
As $\tau$ increases the longitudinal pressure grows and, because the energy-momentum tensor is traceless, the transverse pressure decreases.

\begin{figure}[H]
\centering
\includegraphics[scale=0.072]{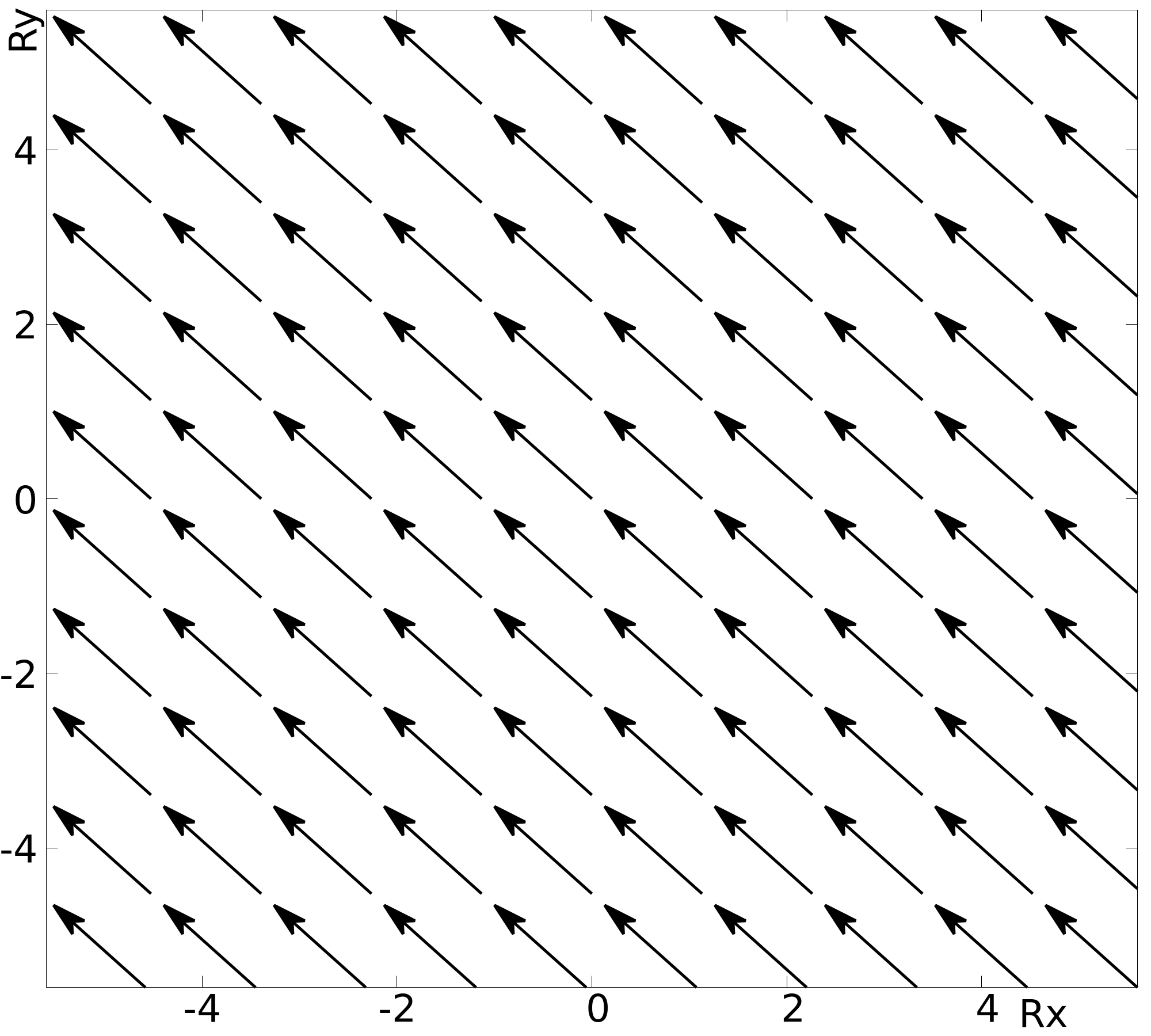} 
\includegraphics[scale=0.072]{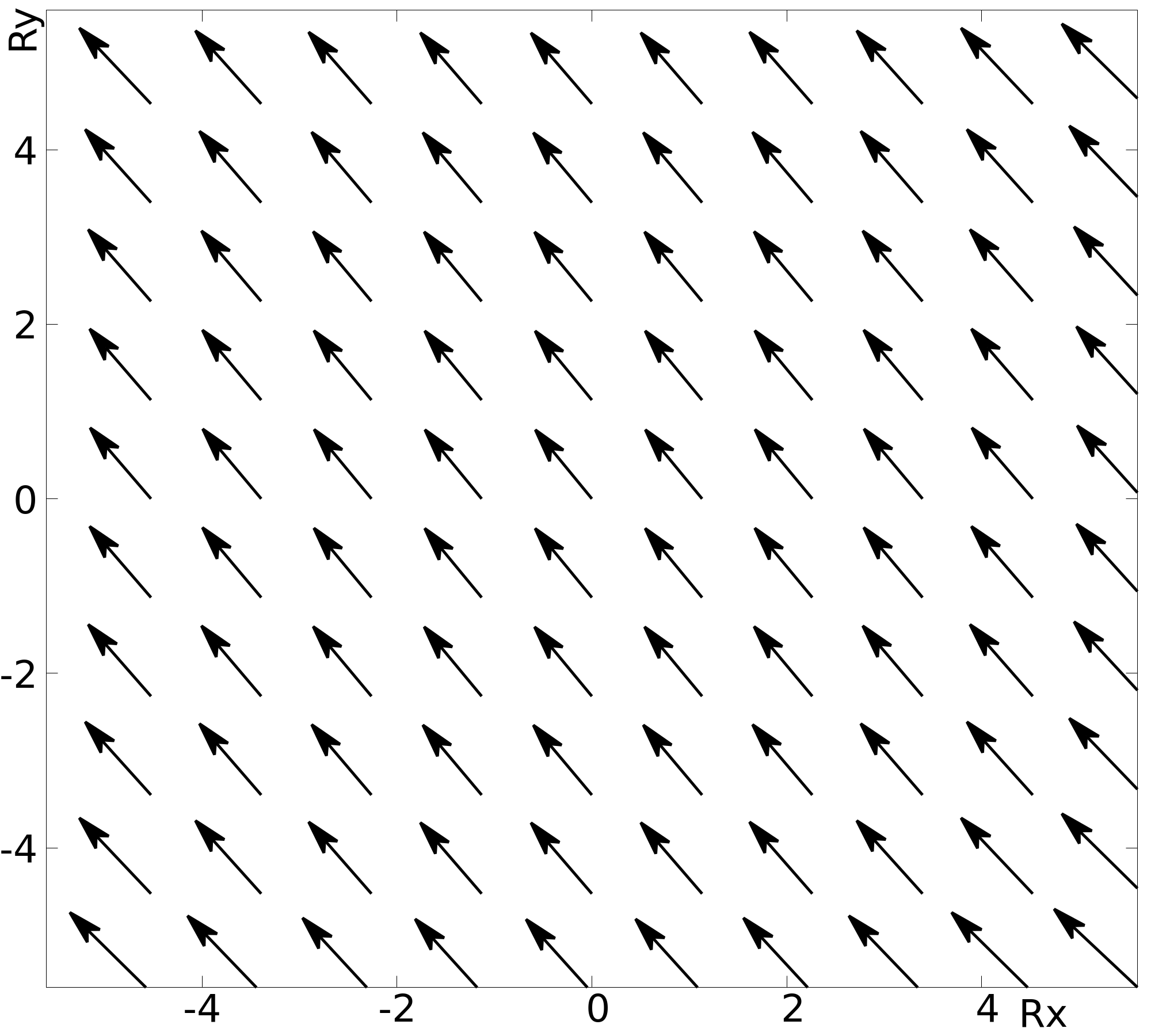} 
\includegraphics[scale=0.072]{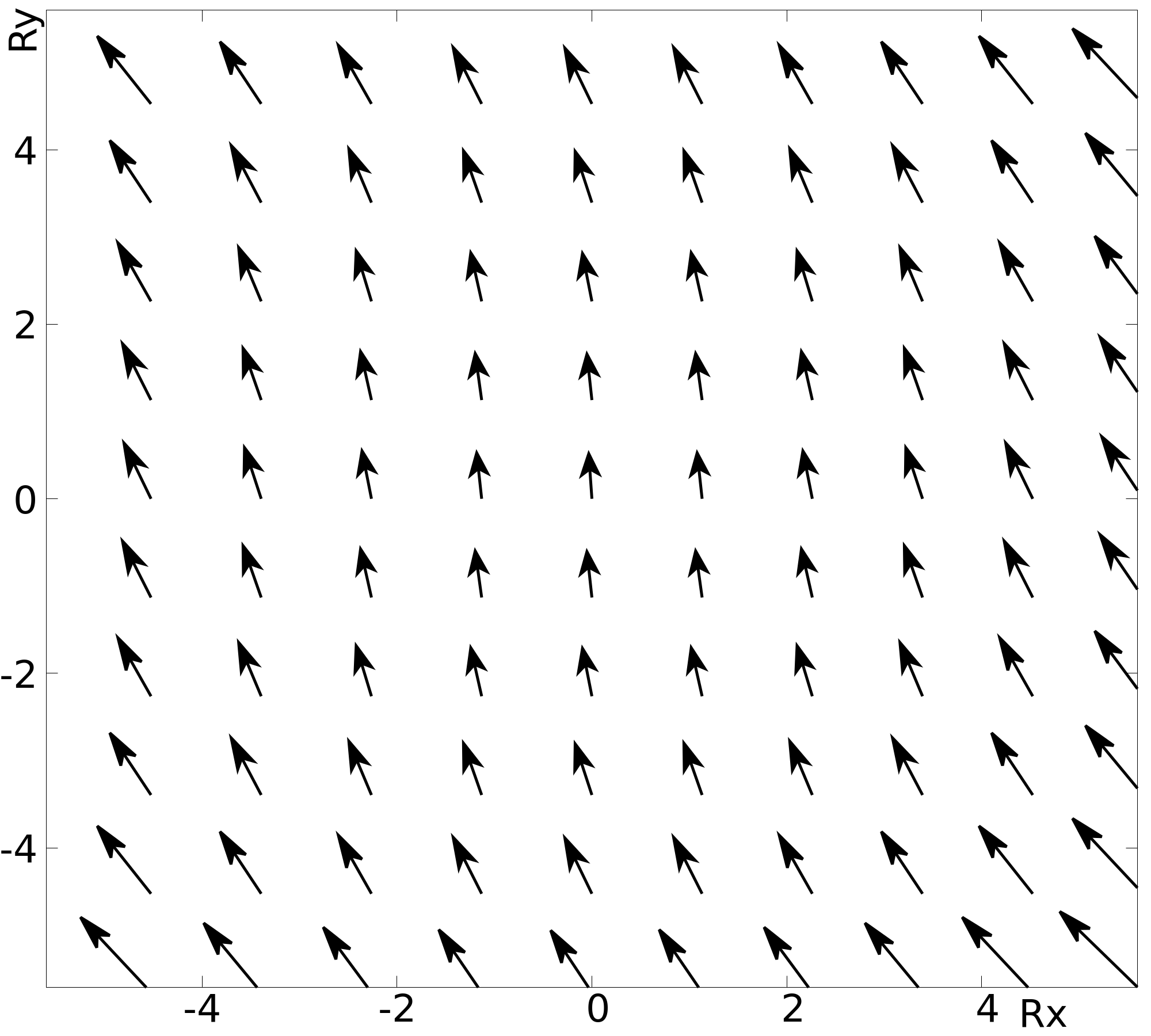} 
\caption{The vector ($p_L/{\cal E}$, $p_T/{\cal E}$)  with $b=\eta=0$ at $\tau=0$ (left), $\tau=0.04$ fm to order $\tau^4$ (middle) and $\tau=0.052$ fm to order $\tau^6$ (right). The axes show $R_x$ and $R_y$ in fm. \label{P-vector}}
\end{figure}

\begin{figure}[H]
\centering
\includegraphics[scale=1.15]{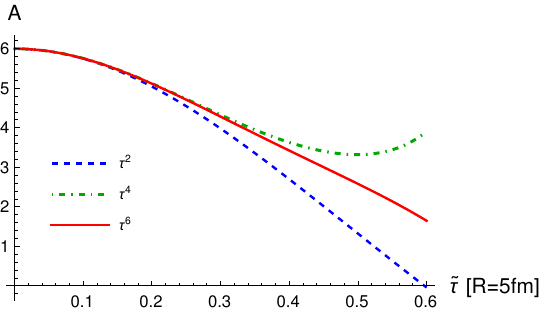} 
\caption{The quantity $A_{TL}$ at $R=5$ fm and $\eta=0$ for three different orders in the $\tau$ expansion.  \label{tau-conv}}
\end{figure}

In figure \ref{P-vector} we show the vector $(p_L/{\cal E}$, $p_T/{\cal E})$ in the transverse plane with $b=\eta=0$. 
In the left panel we see $-p_L=p_T$ at $\tau=0$.
In the next two panels we use the biggest value of $\tau$ for which we trust the $\tau$ expansion at that order (these times are determined from additional results in this and the following sections, see for example figures \ref{tau-conv} and \ref{plotV1V5proj}). 
In the middle figure we include terms to order $\tau^4$ and set $\tau=0.04$ fm. 
The figure shows that the vector has straightened slightly across the transverse plane. 
In the right figure we include all terms to order $\tau^6$ and use $\tau=0.052$ fm. 
We see that the vector has straightened even more, but not uniformly. 

The authors of Ref. \cite{Jankowski:2020itt} suggested that the  anisotropy of the transverse and longitudinal pressures should be characterized using the quantity
\bea
A_{TL} \equiv \frac{3(p_T-p_L)}{2p_T+p_L}\,\label{AA-def}
\eea
which takes the value $A_{TL}=6$ at $\tau=0$ (using equation (\ref{diag})) and would be zero in an equilibrated isotropic plasma. 
We expect that $A_{TL}$ should decrease as $\tau$ increases, up until the point at which the proper time expansion breaks down. This behaviour is observed in
figure \ref{tau-conv} which shows $A_{TL}$ versus $\tilde\tau$ at three different orders in the $\tau$ expansion. The three curves are very close to each other up to about $\tilde\tau=0.2$, and the fourth- and sixth-order expansions agree well up to about $\tilde\tau=0.4$. In section \ref{radial-flow-sec} we show additional evidence that the sixth-order expansion can be trusted to about $\tilde\tau=0.5$. We note that for the simpler case of nuclei 
that are uniform in the transverse plane, resummations of selected sets of terms have been used to increase the range of convergence of the energy and pressure \cite{Carrington:2020ssh,Fukushima:2007yk,Li:2016eqr}.

\begin{figure}[H]
\centering
\includegraphics[scale=1.25]{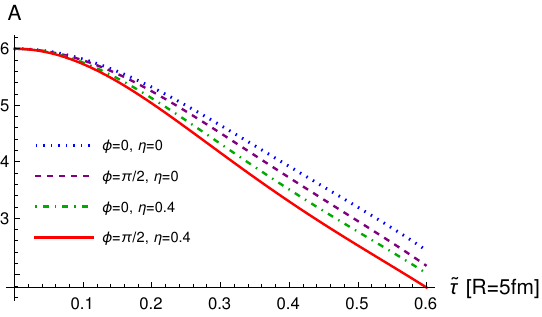} \quad
\caption{The quantity $A_{TL}$ defined in equation (\ref{AA-def}) at $R=5$ fm and $b=6$ fm.  \label{MartB}}
\end{figure}

We  also study how the behaviour of $A_{TL}$ depends on azimuthal angle (denoted $\phi$), spatial rapidity, and impact parameter. 
As expected, $A_{TL}$ moves towards the equilibrium value more quickly when the impact parameter is smaller, and the region where the two ions overlap is greater. 
In figure  \ref{MartB} we show the quantity $A_{TL}$ at sixth order in the proper time expansion  as a function of $\tilde\tau$, for different values of $\eta$ and $\phi$.  
We consider $\phi=0$, which corresponds to $\vec R$ in the reaction plane, and $\phi=\pi/2$, where $\vec R$ is perpendicular to the reaction plane. 
The graph shows that $A_{TL}$ drops more quickly when either the azimuthal angle or the spatial rapidity increases. 

\begin{figure}[H]
\centering
\includegraphics[scale=1.02]{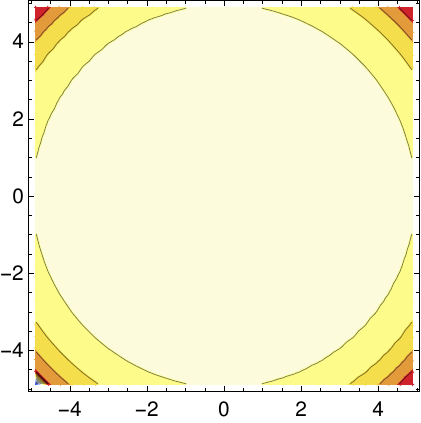} 
\includegraphics[scale=1.0]{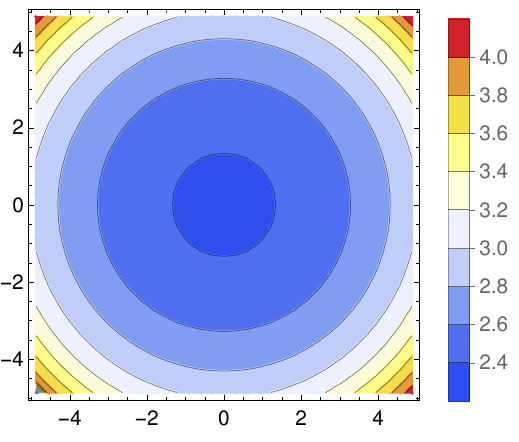} 
\caption{The quantity $A_{TL}$ in equation (\ref{AA-def}) with $b=0$ and $\eta=0$ at $\tau = 0.045$ fm at order $\tau^4$ (left panel) and $\tau^6$ (right panel). 
The axes show $R_x$ and $R_y$ in fm.  \label{more-A}}
\end{figure}

In figure \ref{more-A} we show contour plots of $A_{TL}$ in the transverse plane for $\eta=0$, $b=0$ and $\tau=0.045$ fm at order $\tau^4$ and $\tau^6$.
One sees that when the order $\tau^6$ terms are included, the region of the transverse plane where $A_{TL}$ is small is significantly broader. 

\begin{figure}[H]
\centering
\includegraphics[scale=0.99]{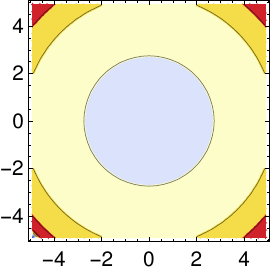} 
\includegraphics[scale=0.99]{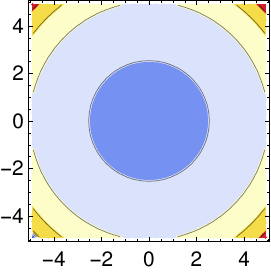} 
\includegraphics[scale=0.99]{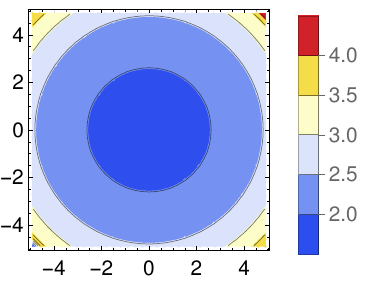} 
\caption{The quantity $A_{TL}$ in equation (\ref{AA-def}) with $b=0$ and $\eta=0$ at order $\tau^6$ for $\tau = 0.04$ fm (left panel), $\tau = 0.045$ fm (centre panel) and $\tau = 0.05$ fm (right panel).  
The axes show $R_x$ and $R_y$ in fm.  \label{e-more-A}}
\end{figure}

In figure \ref{e-more-A} we show contour plots of $A_{TL}$ in the transverse plane for $\eta=0$, $b=0$ at order $\tau^6$ for three different times. One sees that the value of  $A_{TL}$ shrinks across the transverse plane as $\tau$ increases. 

\begin{figure}[H]
\centering
\includegraphics[scale=0.90]{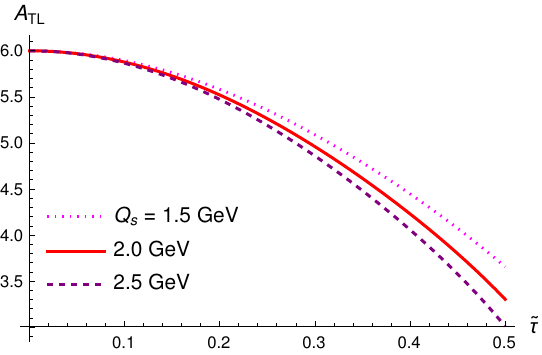} 
\includegraphics[scale=0.88]{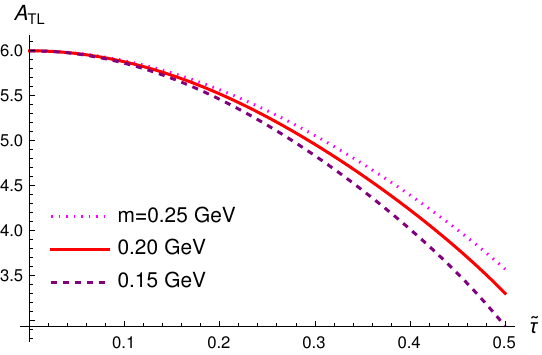}
\caption{The quantity $A_{TL}$ with $R=5$ fm, $b=0$ and $\eta=0$ at order $\tau^6$ for three different values of the saturation scale (left panel) and mass parameter (right panel). \label{Qs_mass}}
\end{figure}

We can also use the quantity $A_{TL}$ to demonstrate that our results are not strongly dependent on the UV and IR scales that enter the calculation ($Q_s$ and $m$ in our notation). This is important because the exact values of these scales are not known, and also because the way that these scales enter the calculation depends on the method chosen to perform the regularization. In all calculations in this paper we have used $Q_s=2.0$ GeV and $m=0.2$ GeV.  
In figure \ref{Qs_mass} we show $A_{TL}$ at order $\tau^6$ as a function of time for three different values of $Q_s$ with $m=0.2$ GeV (left panel) and for three different values of $m$ with $Q_s=2.0$ GeV (right panel).
The graphs show that within the range of validity of the $\tau$ expansion, the dependence on the value of these scales is weak. 

\subsubsection{Transverse pressure anisotropy}
\label{sec-curlya}

In this section we look at a quantity that can be used to characterize the asymmetry of the transverse pressure. We define \cite{Krasnitz:2002ng} 
\bea
\{A_{xy}\}
 \equiv  \frac{\langle T^{yy}-T^{xx}\rangle}{\langle T^{xx}+T^{yy}\rangle}
\label{vareps}
\eea
where the angular brackets indicate integration over the transverse plane. 
For comparison we will also calculate 
\bea
\{ A_{TL} \}  \equiv \frac{3\langle p_T-p_L\rangle}{\langle 2p_T+p_L\rangle }\,.\label{AA-int}
\eea

\begin{figure}[H]
\centering
\includegraphics[scale=1.2]{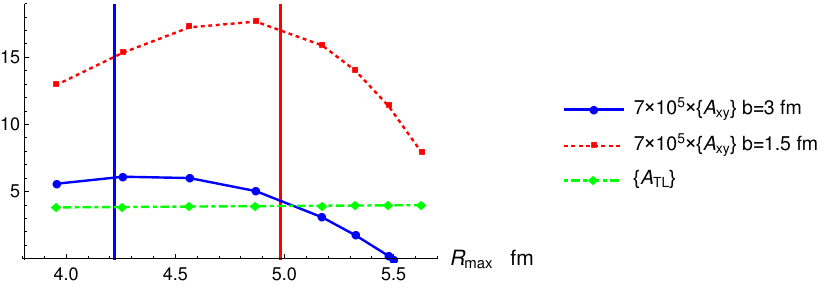} 
\caption{The quantities $\{A_{TL}\}$ and $7 \times 10^5 \times \{A_{xy}\}$  at $\tau=0.04$ fm and $\eta=0$ as functions of $R_{\rm max}$.  \label{Ahat-plot}}
\end{figure}

The leading-order contribution to $\{A_{xy}\}$ comes from the first order term in the gradient expansion and therefore this quantity, in contrast to $\{A_{TL}\}$, will be sensitive to the region of the transverse plane that is close to the edges of the nuclei. 
We must verify that the integral is largely independent of the upper limit that is used to perform the two dimensional integration over the transverse plane, which we call $R_{\rm max}$. 
In figure \ref{Ahat-plot} we show $\{A_{xy}\}$ at $\tau=0.04$ fm for two different values of impact parameter, as a function of $R_{\rm max}$. 
The two vertical lines in the figure represent the values of $R_{\rm max}$, for each value of $b$, for which the parameter $\delta$ in equation (\ref{delta-def}) is less than 0.6 for both nuclei. 
One sees that the result for $\{A_{xy}\}$ grows as $R_{\rm max}$ increases, up to approximately the value of $R_{\rm max}$ at which the gradient expansion breaks down. 
Smaller impact parameters give results that can be trusted up to larger values of $R_{\rm max}$, as explained in section \ref{sec-gradient}. 
On the same graph we show the result for $\{A_{TL}\}$ at $\tau=0.04$ fm and $\eta=0$. The change with impact parameter is too small to be seen on the graph, and the result is almost six orders of magnitude larger than $\{A_{xy}\}$ and nearly completely independent of $R_{\rm max}$. 
Figure \ref{Ahat-plot} clearly illustrates the problem discussed in section \ref{sec-gradient}. 
The nature of the gradient expansion makes it difficult to calculate any quantity that gets important contributions from the region of the transverse plane that corresponds to the edges of the nuclei. We will return to this point in sections \ref{sec-f-coefficients} - \ref{sec-L2}.

\subsection{Radial flow}
\label{radial-flow-sec}

\begin{figure}[H]
\centering
\includegraphics[scale=0.85]{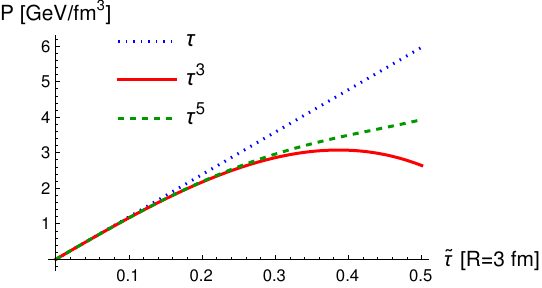} \quad 
\includegraphics[scale=0.85]{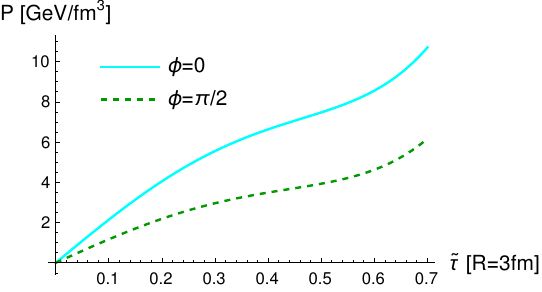} \quad
\caption{The radial flow at $\eta=0$, $b=6$ fm and $R=3$ fm. The left panel shows different orders in the proper time expansion at $\phi=\pi/2$, and the right panel shows two different values of the azimuthal angle at order $\tau^5$.   \label{plotV1V5proj}}
\end{figure}

To describe the radial flow of the expanding glasma in the transverse plane we look at the radial projection of the transverse Poynting vector $P \equiv \hat R^i T^{i0}$.
In the left panel of figure \ref{plotV1V5proj} we show this quantity at $\eta=0$, $b=6$ fm, $R=5$ fm and $\phi=\pi/2$ at different orders in the $\tau$ expansion. 
At lowest order $P$ increases linearly with  time. Including higher order contributions we see that $P$ increases more slowly with time as the system expands. If we keep only terms at order $\tau^3$, it appears that $P$ actually starts to decrease when $\tilde\tau \gtrsim 0.4$, however the near field expansion is not valid for these times when only up to cubic terms are included. The result at order $\tau^5$ shows a less pronounced flattening up to about $\tilde\tau=0.5$, which again indicates the limit of validity of the near field expansion. In the right panel of figure \ref{plotV1V5proj} we show $P$ at $\eta=0$, $b=6$ fm and $R=5$ fm for $\phi=0$, which corresponds to $\vec R$ in the reaction plane, and $\phi=\pi/2$, where $\vec R$ is perpendicular to the reaction plane. One sees that the flattening is more pronounced when the azimuthal angle is smaller. 

\begin{figure}[H]
\centering
\includegraphics[scale=0.0756]{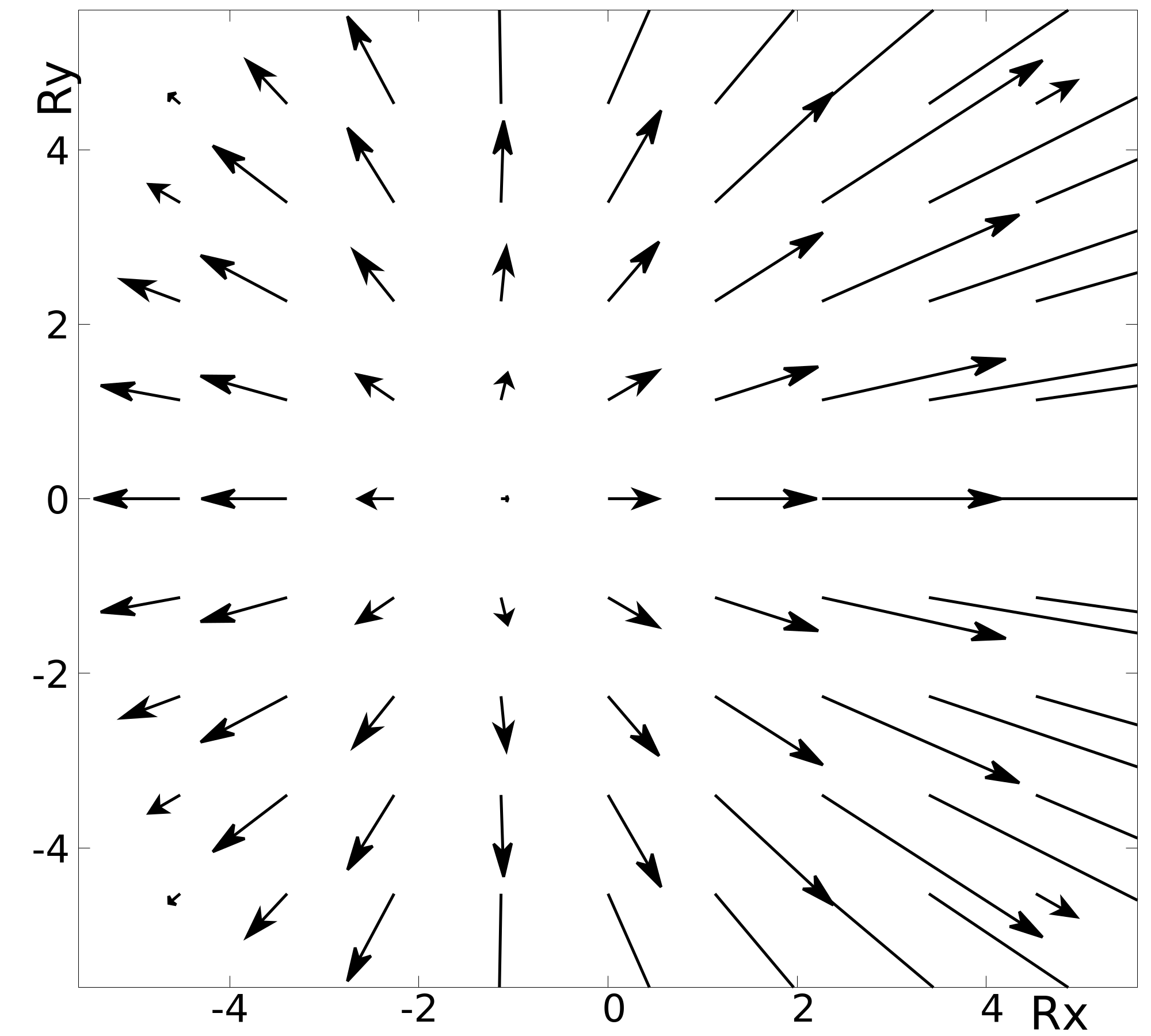} 
\includegraphics[scale=0.0756]{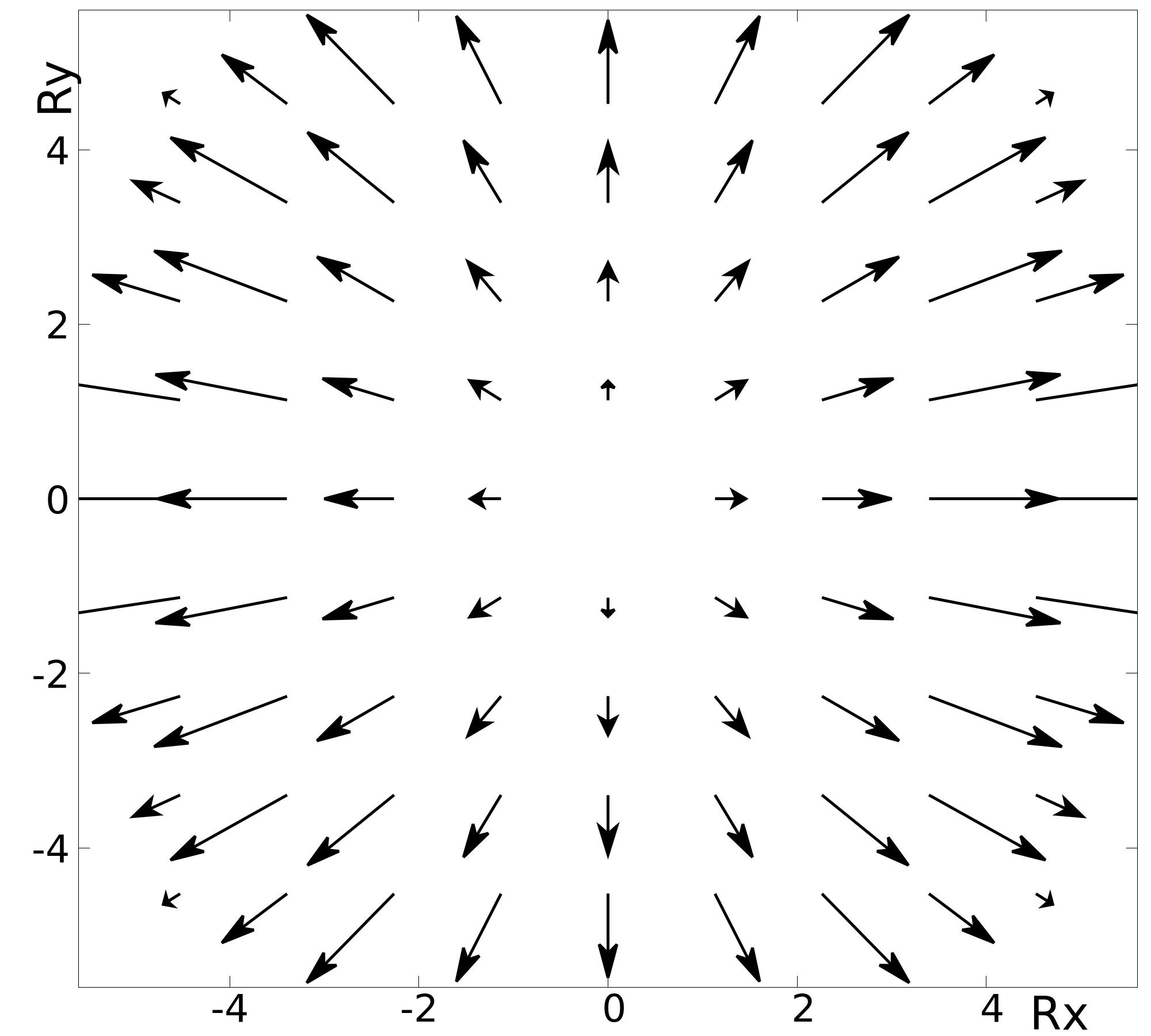} 
\includegraphics[scale=0.0756]{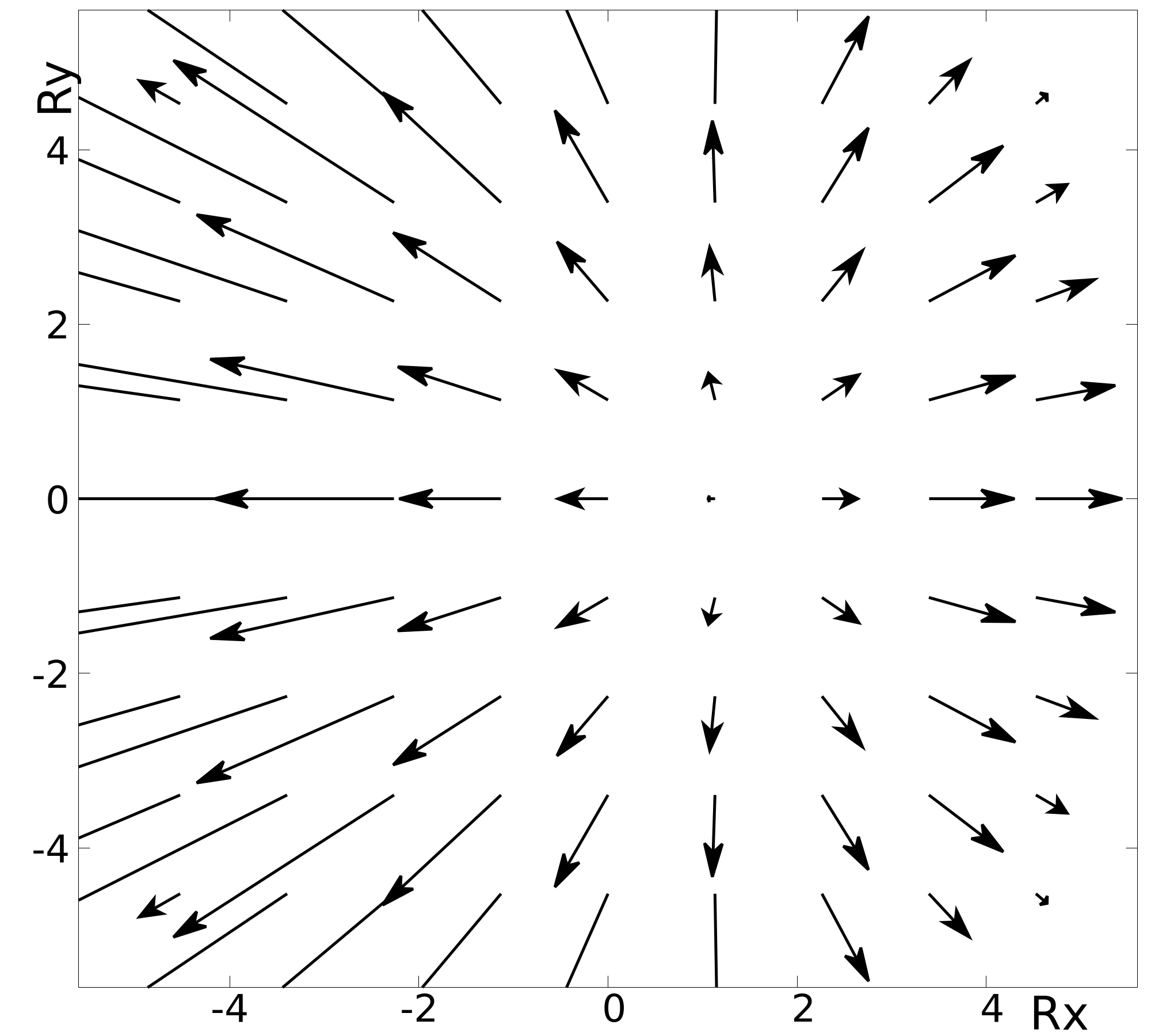} 
\caption{The Poynting vector in the transverse plane with $b=6$ fm and $\tau=0.05$ fm for $\eta=-1.0$ (left), $\eta=0$ (centre), and $\eta=1.0$ (right).  
   \label{Ti0-trans}}
\end{figure}

\begin{figure}[H]
\centering
\includegraphics[scale=0.1096]{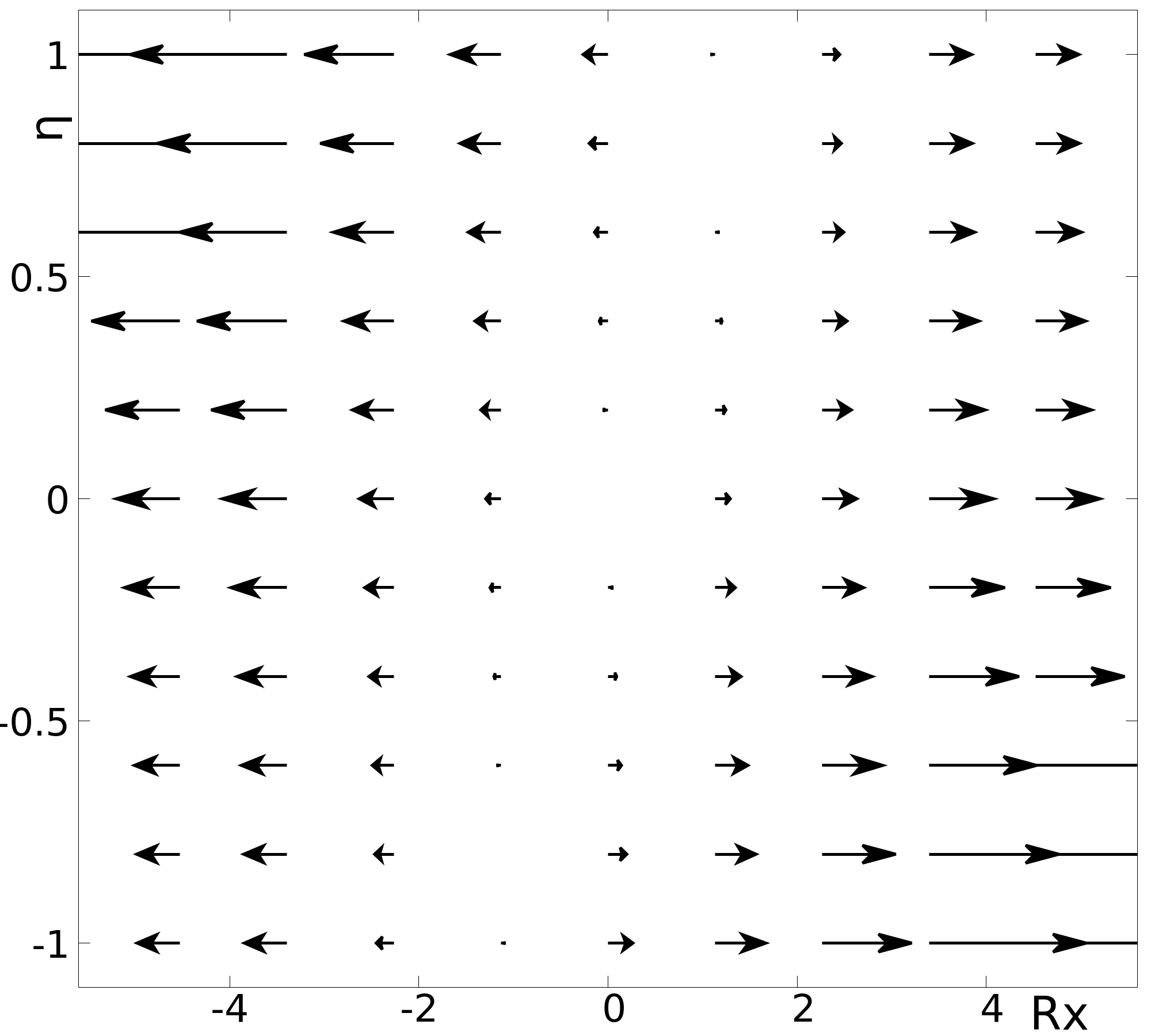} 
\includegraphics[scale=0.1096]{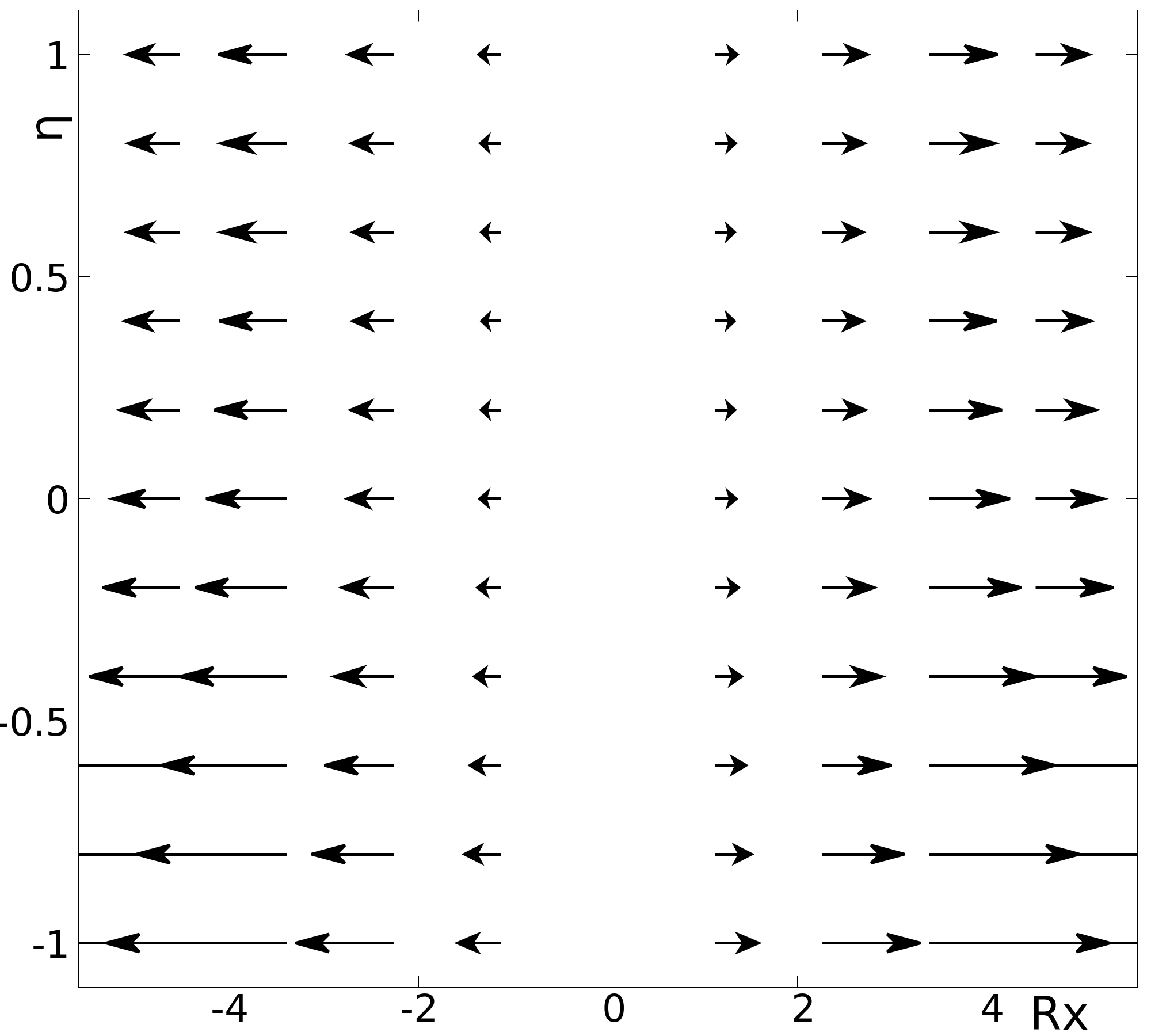} 
\caption{The components of the Poynting vector at $\tau=0.05$ fm with $R_y=0$ and $A_1=207$ for $b=6$ fm and $A_2=207$ (left panel), and for $b=0$ and $A_2=40$ (right panel). The axes show $R_x$ in fm and $\eta$.
   \label{Ti0-eta}}
\end{figure}

\begin{figure}[H]
\centering
\includegraphics[scale=0.0999]{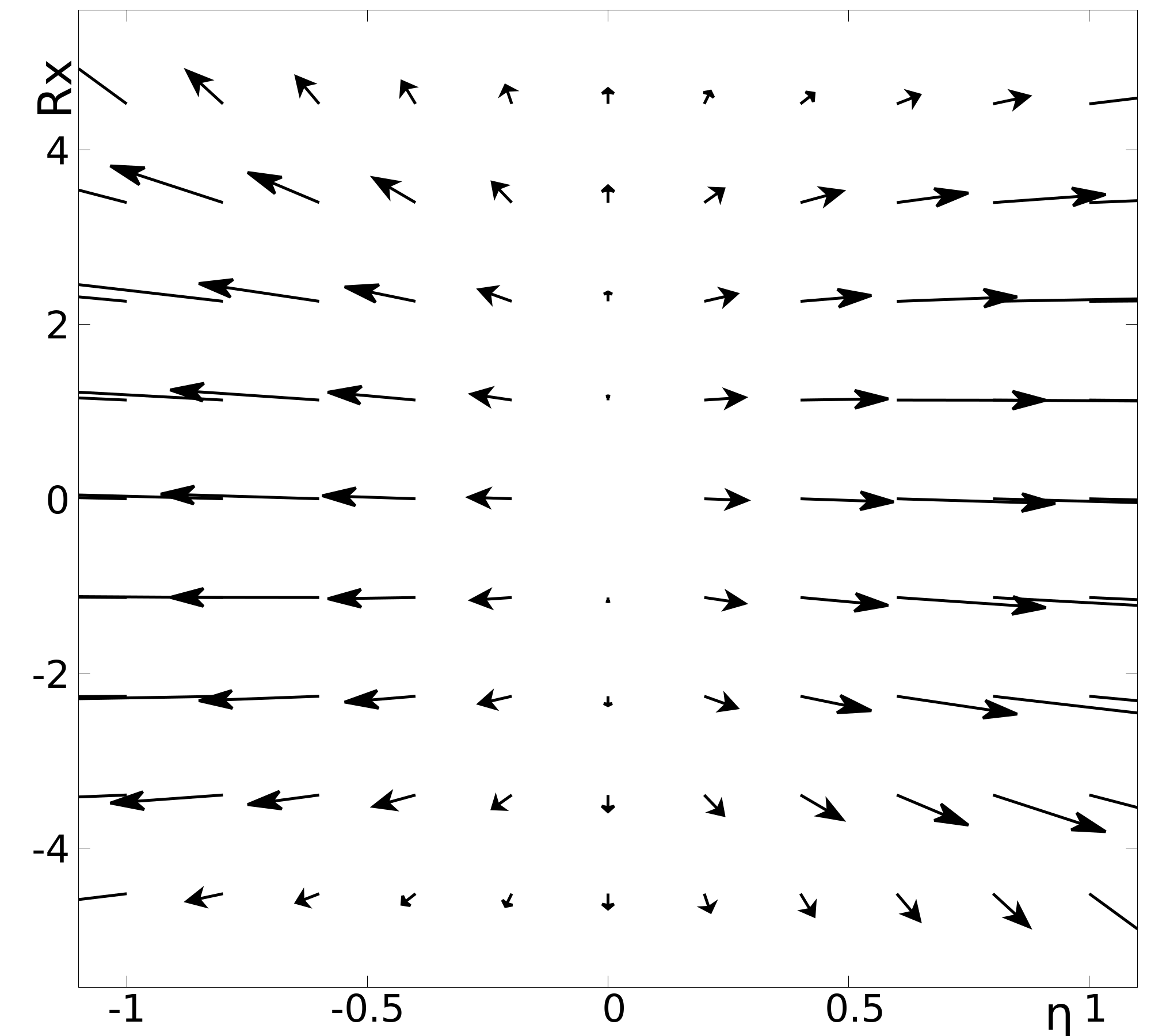} \quad
\includegraphics[scale=0.0999]{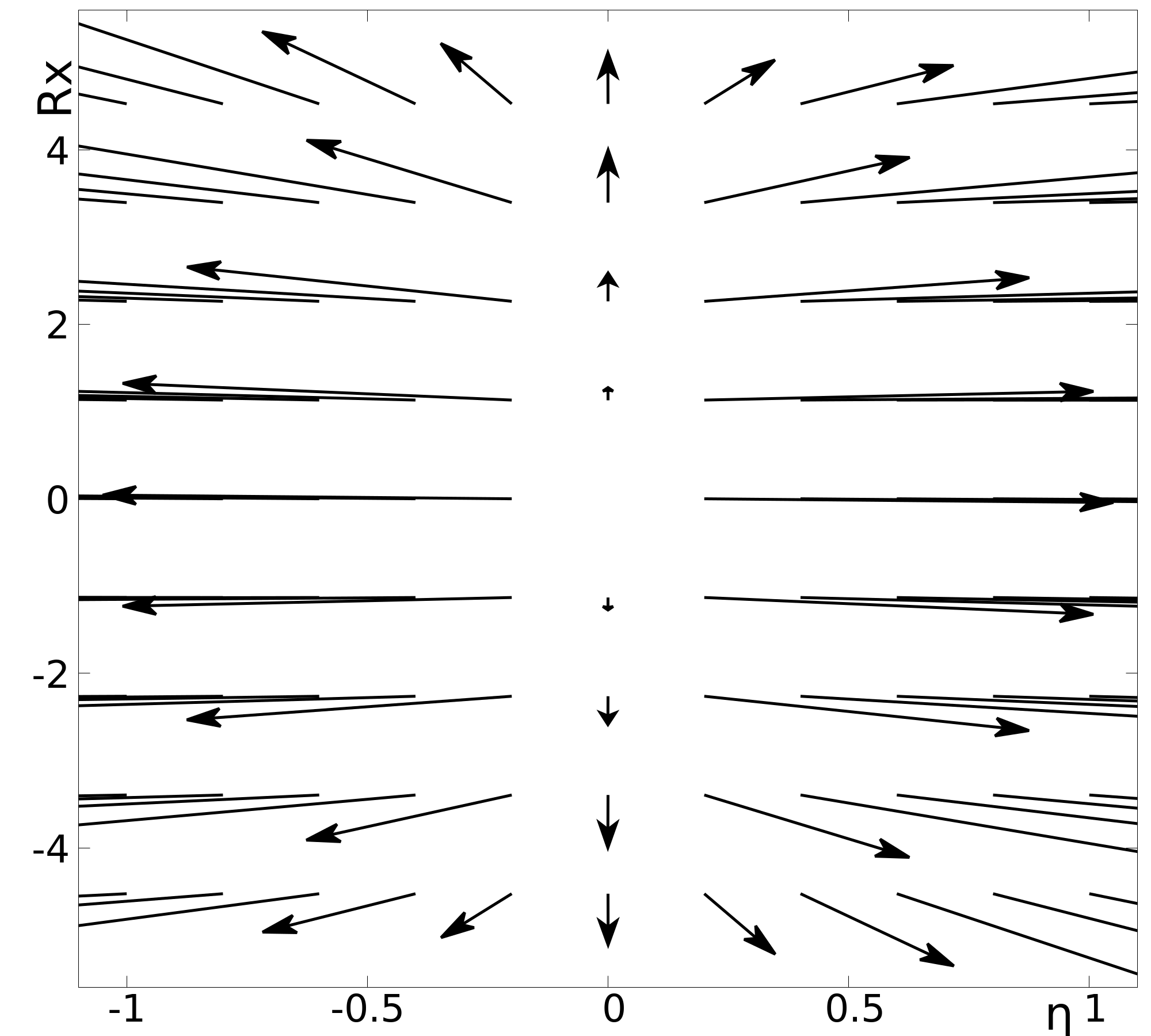} \quad
\includegraphics[scale=0.0999]{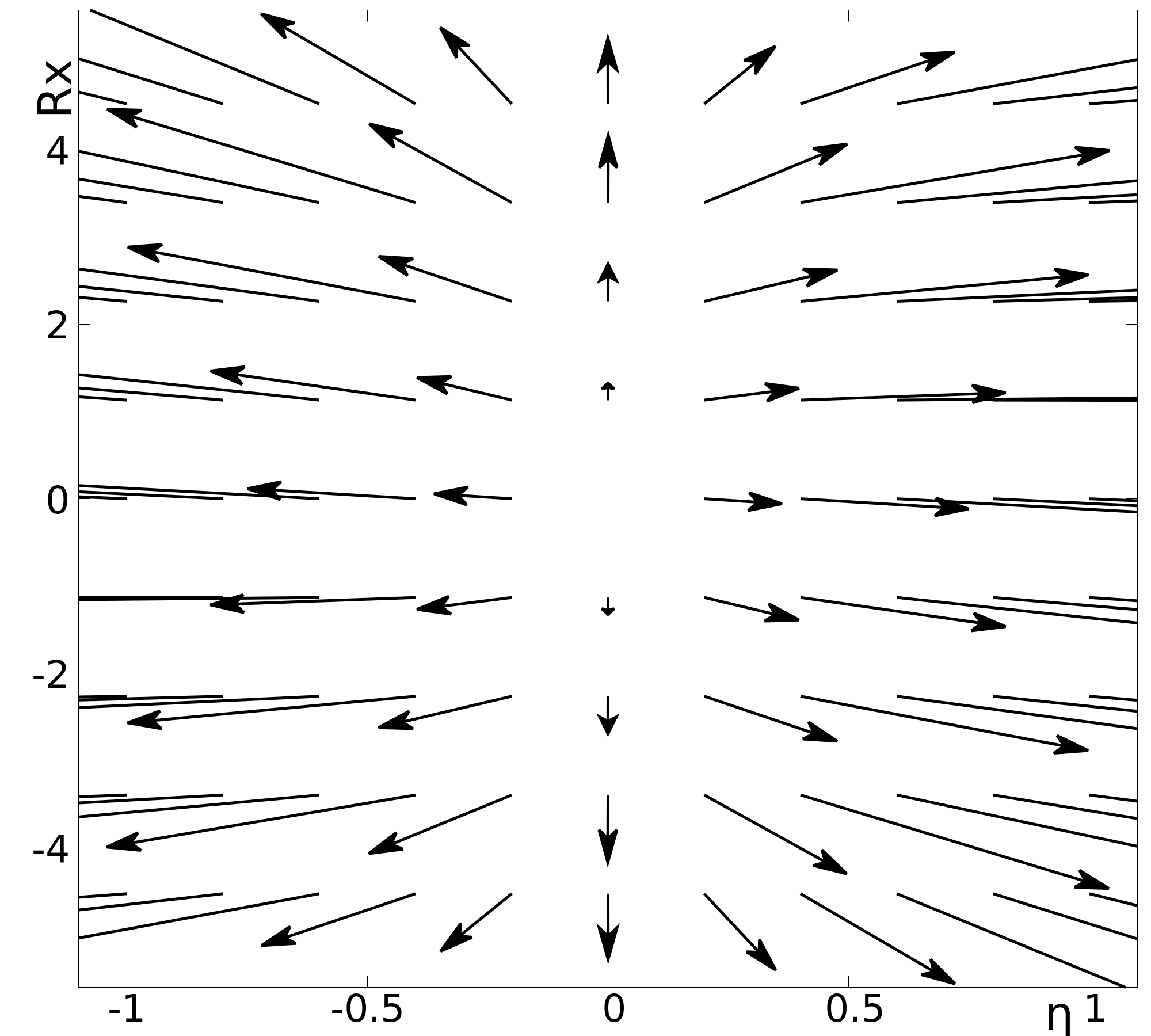} \quad
\includegraphics[scale=0.0999]{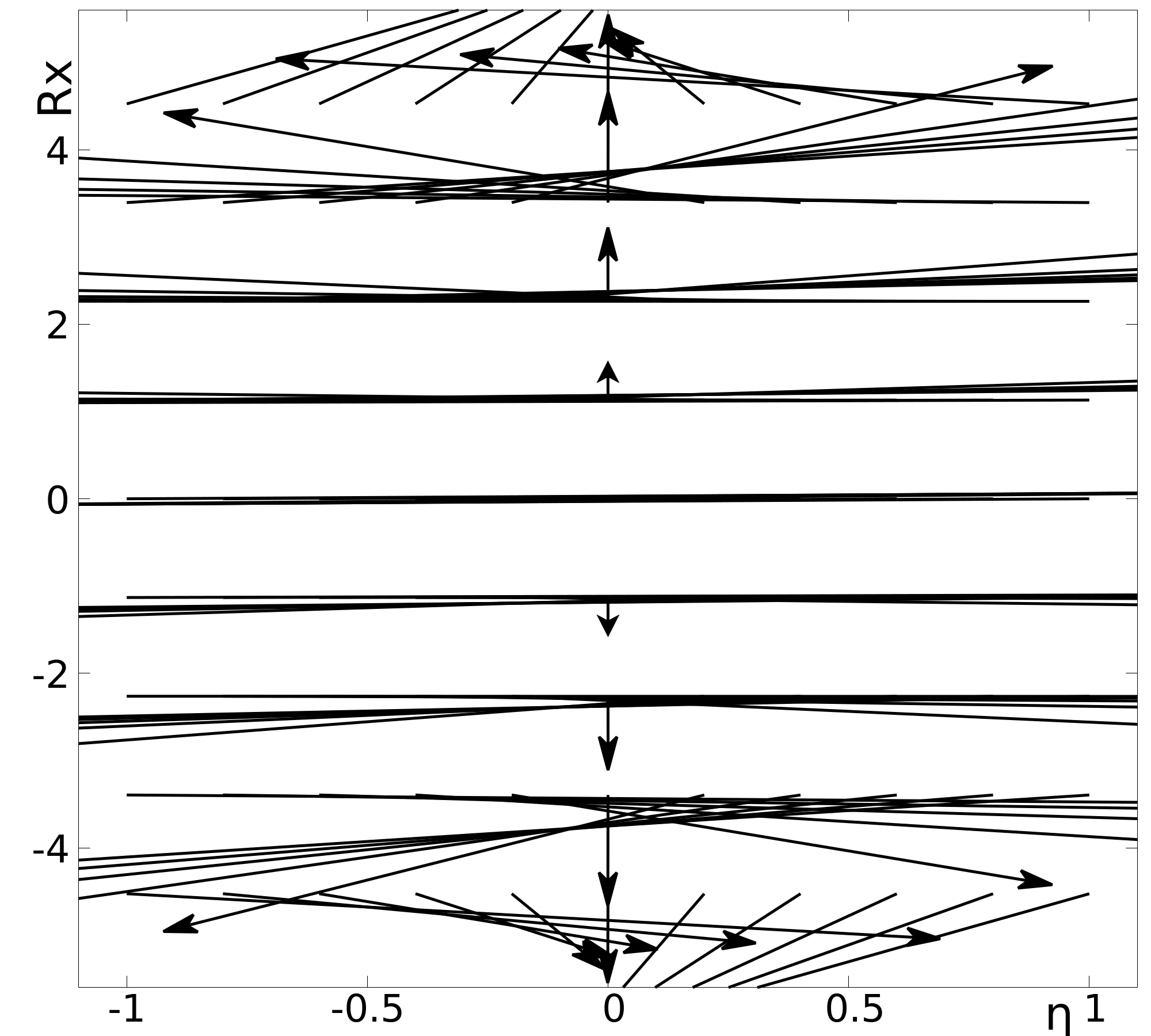} \quad
\caption{The vector field ($T^{z0}$, $10\,T^{x0}$) with $b=6$ fm and $R_y=0$. 
The times are $\tilde\tau=0.1$ (top left), $\tilde\tau=0.4$ (top right), $\tilde\tau=0.6$ (bottom left) and $\tilde\tau=0.8$ (bottom right). At early times the motion is predominantly longitudinal, and transverse velocity components develop at later times. 
The last figure is beyond the time at which we trust the near field expansion. 
   \label{vzvx}}
\end{figure}

Figures \ref{Ti0-trans}-\ref{vzvx} show the Poynting vector in arbitrary units in the transverse plane for $b=6$ fm and $\tau=0.05$ fm. 
The three panels in figure \ref{Ti0-trans} correspond to three different values of fixed rapidity. One sees that for positive/negative values of rapidity, the centre of the collision moves towards the ion moving in the positive/negative $z$-direction which has been displaced in the positive/negative $x$-direction. 
The positions where the magnitude of the Poynting vector is zero for the three cases shown are: $(R_x,R_y)=(0,-0.91)$ fm for $\eta=-1.0$, $(0,0)$ for $\eta=0$ and $(0,0.91)$ fm for $\eta=1.0$.

In figure \ref{Ti0-eta} we show the Poynting vector at $R_y=0$ in the ($\eta$,$R_x$) plane, in arbitrary units.  
In the first panel we see that when $\eta$ is positive, which corresponds to a position closer to the right moving ion that has been displaced in the positive $x$ direction, the sign of the Poynting vector is predominantly negative. This corresponds to the negative value of the directed flow coefficient discussed in section \ref{sec-f-coefficients}. The second panel shows that the system expands more strongly in the wake of the larger nucleus. 

In figure \ref{vzvx} we show the vector field ($T^{z0}$, $10\,T^{x0}$) with $b=6$ fm and $R_y=0$, in arbitrary units. One sees the transverse velocity components develop up until the time at which the $\tau$ expansion breaks down. 

\subsection{Fourier coefficients of the azimuthal distribution}
\label{sec-f-coefficients}

In this section  we calculate Fourier coefficients of the azimuthal distribution of the flow vector $T^{i0}(\vec x_\perp)$. 
The flow vectors are used as input in hydrodynamic codes, and the Fourier coefficients are  related to experimental observables. 
In Appendix \ref{sec-azi-distro} we define the form of the azimuthal distribution that we use and explain how the Fourier coefficients $v_n$ are calculated. 

We note that one of the important sources of uncertainty in experimentally determined values of Fourier coefficients are fluctuations in the positions of the nucleons that directly participate in the primary interaction, which produce deviations between the orientation of the event-plane, determined by the principal axes of the participants, and the reaction plane. In our calculation the reaction plane is known, and fluctuations are not included. 
One consequence is that the Fourier coefficients in our calculation exhibit specific symmetries: the coefficients $v_n$ with $n$ odd are rapidity odd, and those with $n$ even are rapidity even. 

\begin{figure}[]
\includegraphics[scale=0.56]{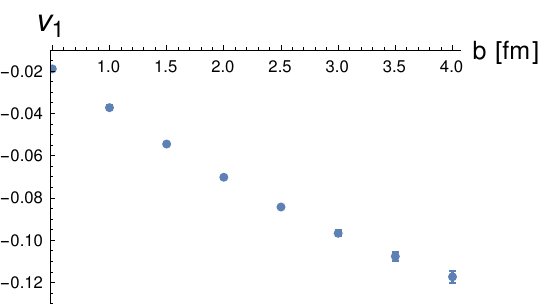} \;
\includegraphics[scale=0.56]{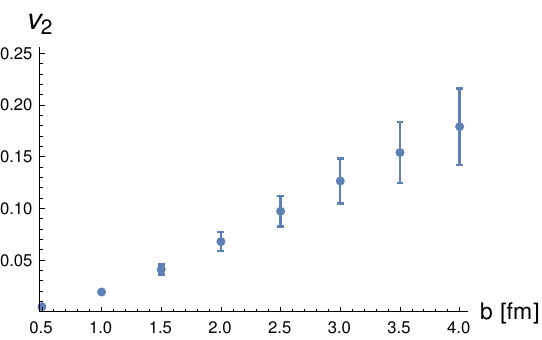} \;
\includegraphics[scale=0.58]{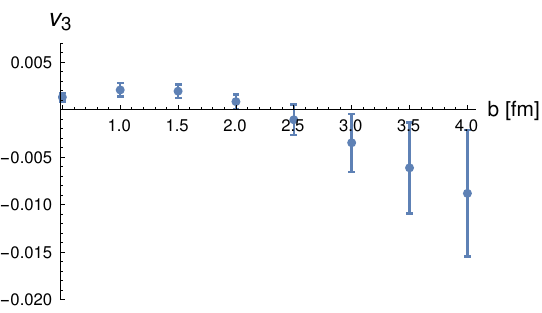} 
\caption{The Fourier coefficients $v_1$, $v_2$ and $v_3$ with $\eta=0.5$ and $\tau=0.04$ fm as functions of impact parameter. The procedure for calculating the error bars is explained in the text.  \label{plot-ebars} }
\end{figure}

To estimate the error in the Fourier coefficients, we perform the integrals using, for each value of $b$, a set of 15 evenly spaced values of $R_{\rm max}$ between 4.5 fm and 7.0 fm. 
We average the results and calculate the error bar from the standard deviation. 
In figure \ref{plot-ebars} we show the first three Fourier coefficients at $\eta=0.5$ and $\tau=0.04$ fm. As in section \ref{sec-curlya}, we find that the calculation is reasonably insensitive to the upper limit of the integration for impact parameters $b \lesssim 2.5$ fm. 

\begin{figure}[H]
\includegraphics[scale=0.90]{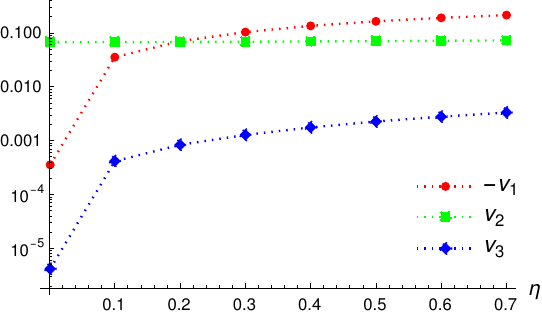} \quad
\includegraphics[scale=0.90]{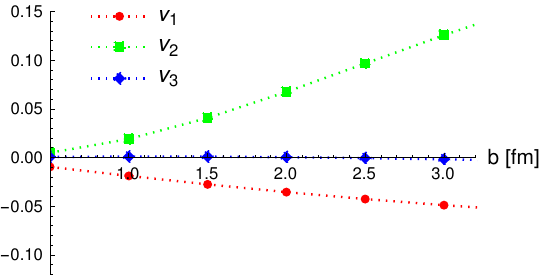} \quad
\caption{The Fourier coefficients $v_1$, $v_2$ and $v_3$ at $\tau=0.04$ fm. 
The left panel is for fixed $b=2$ fm and the right panel is fixed $\eta=0.1$.
\label{plotv12} }
\end{figure}

\begin{figure}[H]
\includegraphics[scale=0.56]{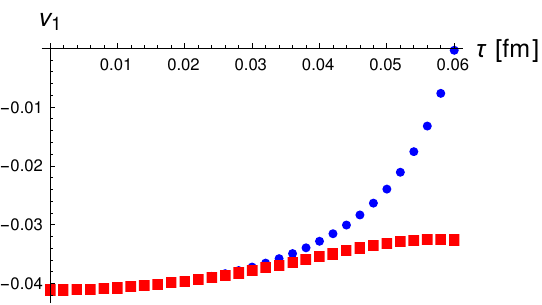} \,
\includegraphics[scale=0.56]{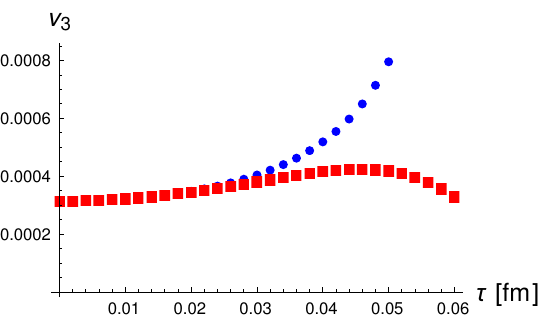} \,
\includegraphics[scale=0.58]{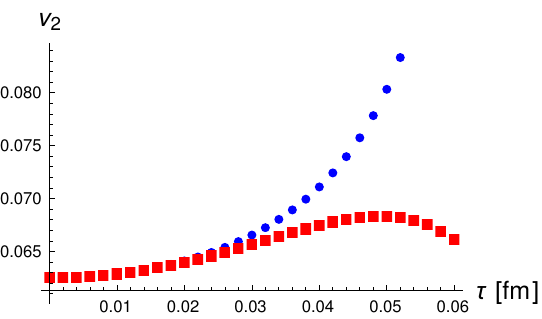} \,
\caption{The Fourier coefficients $v_1$, $v_2$ and $v_3$ at $\eta=0.1$ and $b=2$ fm as functions of time. The blue dots show the result at order $\tau^3$ and the red squares represent the order $\tau^5$ results. 
\label{plotv123time} }
\end{figure}

In figure \ref{plotv12} we look at the first three Fourier coefficients as functions of rapidity and impact parameter, at $\tau=0.04$ fm.
We use $R_{\rm max}=5.9$ fm in all calculations.
In the left panel of figure \ref{plotv12} we show $v_1$, $v_2$ and $v_3$ at $\tau=0.04$ fm with $b=2$ fm as a function of rapidity. 
In the right panel we show the same three Fourier coefficients at $\tau=0.04$ fm and $\eta=0.1$ as functions of impact parameter. 
The directed flow coefficient, $v_1$, is negative (for $\eta>0$), and the elliptic flow coefficient, $v_2$, is positive.  
The triangle coefficient $v_3$ is small and positive (for $\eta>0$) when $b\lesssim$ 2 fm. 
In figure \ref{plotv123time} we show $v_1$, $v_2$ and $v_3$ as functions of time with $\eta=0.1$ and $b=2$ fm, at order $\tau^3$ and order $\tau^5$. 
In all cases the curves agree well at small times. The order $\tau^3$ curves bend sharply upward at $\tau\approx 0.03$ fm, 
which shows the breakdown of the $\tau$ expansion at third order. At larger values of $\tau$, the order $\tau^5$ results show the opposite behaviour, bending rapidly downward.  
Similar behaviour is seen in figure \ref{tau-conv}, and is discussed in section \ref{sec-martinez}. We  determine numerically that the position of the peak for each of the three curves that show order $\tau^5$ results is $\tau \approx 0.05$ fm, which is approximately the time at which the expansion breaks down. The time for which the second derivative is zero, which corresponds to the point where the curves start to flatten, is approximately $\tau\approx 0.03$ fm for all three curves. This flattening of all three curves inside the region where the $\tau$ expansion converges provides some evidence that the Fourier coefficients will not change rapidly immediately outside the region of validity of the near field expansion. 
We also comment that the radial flow shows similar behaviour (see figure \ref{plotV1V5proj}). 

Our results for the second and third Fourier coefficients are of the same order as experimental values \cite{ATLAS:2012at,Heinz:2013th}, although our result for $|v_1|$ is much bigger than expected \cite{Selyuzhenkov_2011}. 
However, it is usually assumed that anisotropy develops mostly during the hydrodynamic evolution that follows the glasma phase, 
and in this context our results are surprisingly large for all three Fourier coefficients.
Azimuthal correlations in glasma have been investigated by several other groups, using different approaches and looking at different regimes, which makes comparison difficult. 
In Ref. \cite{Krasnitz:2002ng} the authors use a different method and consider impact parameters that are larger than the maximum value for which the gradient expansion can be trusted, and times that are beyond the validity of the near field expansion, but still obtain smaller values of $v_2$. 
In Ref. \cite{Schenke:2015aqa, Lappi:2015vta} the origins of azimuthal correlations in the CGC approach are studied, but quantitative results for a glasma system are not obtained. 

\subsection{Eccentricity}
\label{sec-ecc}

\begin{figure}[H]
\centering
\includegraphics[scale=0.86]{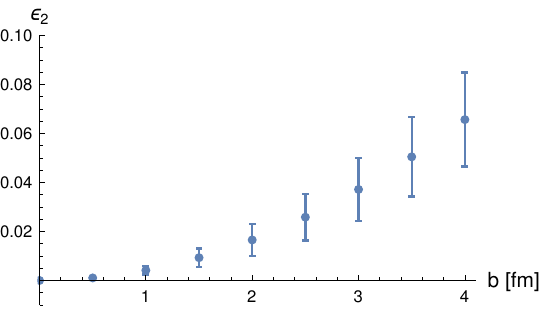}
\caption{The eccentricity $\varepsilon_2$  as a function of impact parameter.  \label{plotecX}}
\end{figure}

\begin{figure}[H]
\centering
\includegraphics[scale=0.82]{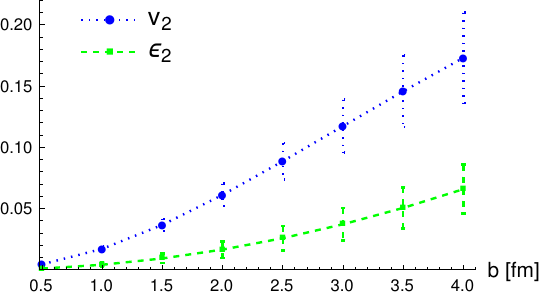}\quad
\includegraphics[scale=0.82]{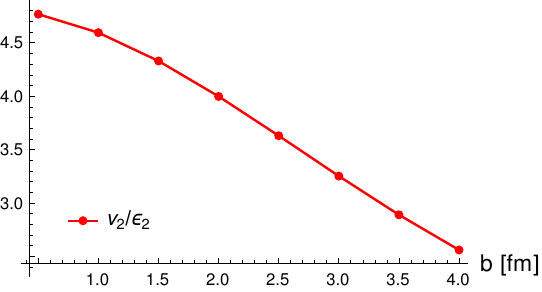}
\caption{The left panel shows the  elliptic flow coefficient $v_2$ and the eccentricity $\varepsilon_2$, and the right panel shows their ratio. All results are at $\tau=0.04$ fm and $\eta=0$.   \label{plotecY}}
\end{figure}

\begin{figure}[H]
\centering
\includegraphics[scale=0.96]{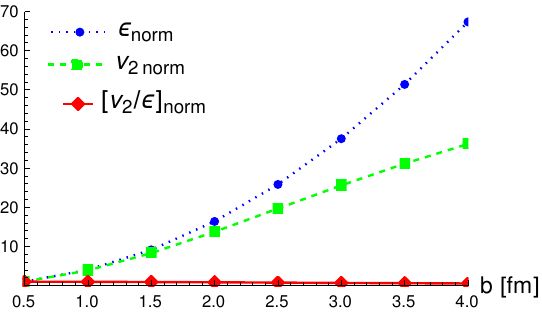}
\caption{The eccentricity $\varepsilon_2$, the elliptic flow coefficient $v_2$, and their ratio at $\tau=0.04$ fm and $\eta=0$ with all curves normalized so that the value at $b=0.5$ fm is set to one.  \label{plotecYb}}
\end{figure}

Spatial deviations from azimuthal symmetry can be characterized with the quantity \cite{Blaizot:2014nia,Demirci:2021kya}
\bea
{\varepsilon}_n = -\frac{\int d^2\vec R |\vec R| \cos(n \phi) {\cal E}(\vec R)}{\int d^2 \vec R |\vec R|  {\cal E}(\vec R)} \text{~~with~~} \phi=\tan^{-1}(R_y/R_x)\,
\label{ecc-def}
\eea
where ${\cal E}(\vec R)$ denotes the energy density.\footnote{The sign in equation (\ref{ecc-def}) is chosen to agree with equation (44) of Ref. \cite{Voloshin:2008dg}.} 
In figure \ref{plotecX} we show the eccentricity, $\varepsilon_2$, as a function of impact parameter at $\tau=0.04$ fm and $\eta=0$. The error bars are calculated from the standard deviation of the results obtained using 15 values of $R_{\rm max}$ that are evenly spaced between 4.5 fm and 7.0 fm.
The results show that $\varepsilon_2$ is largely insensitive to $R_{\rm max}$ for impact parameters $b\lesssim 2.5$ fm. 
In figure \ref{plotecY} we show $v_2$, $\varepsilon_2$ and $v_2/\varepsilon_2$ at $\tau=0.04$ fm and $\eta=0$ using $R_{\rm max}=5.9$ fm. 
In figure \ref{plotecYb} we show the same curves normalized so that the value at $b=0.5$ fm is set to one. 
The results show a clear correlation between the spatial asymmetry introduced by the initial geometry, and the anisotropy of the azimuthal distribution of the gluon momentum. Correlations of this kind are considered characteristic of the onset of hydrodynamic behaviour. 

\subsection{Angular Momentum}
\label{sec-L2}

\begin{figure}[H]
\centering
\includegraphics[scale=0.75]{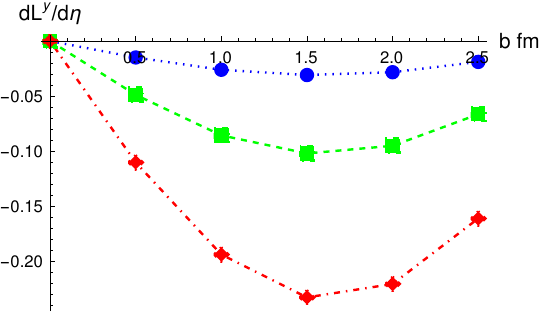} \quad
\includegraphics[scale=0.75]{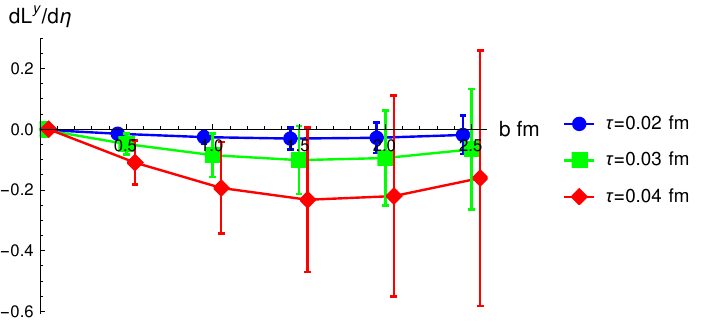} 
\caption{The angular momentum per unit rapidity  as a function of impact parameter. The right panel shows the same three curves as in the left panel but with error bars included, and the values of impact parameter are shifted 0.05 fm left for the $\tau=0.02$ fm curve, and 0.05 fm right for the $\tau=0.04$ fm curve, so that the error bars are separated enough to be seen individually.   \label{plotL1}}
\end{figure}

The angular momentum of the glasma can be calculated from equation (\ref{L-def-4}). 
In the left panel of figure \ref{plotL1} we show our results at three different times, using $R_{\rm max} = 5.9$ fm. 
The error bars in the right panel are obtained by calculating the standard deviation of the results using a set of evenly spaced values for $R_{\rm max}$ between 4.5 fm and 7.0 fm. 
The error bars that are produced by this procedure are large, even for small impact parameters (for comparison see figures \ref{plot-ebars} and \ref{plotecX}).
This happens because the problem discussed in section \ref{sec-gradient} is much more serious in the calculation of angular momentum than it was in sections \ref{sec-f-coefficients} and \ref{sec-ecc}. The dominant contribution to the angular momentum comes from the parts of the nuclei that are farthest from the collision centre, with respect to which angular momentum is calculated, but these are the regions where the gradient expansion is least to be trusted. 
The angular momentum is negative, which is expected when the ions moving in the positive/negative $z$ directions are displaced in the positive/negative $x$ directions.  
We note that in spite of the size of the error bars in figure \ref{plotL1}, the general shape of the curves is consistently reproduced when the value of $R_{\rm max}$ is changed.
Furthermore, this shape matches the basic form of the results in Refs. \cite{Gao:2007bc, Becattini:2007sr}. 
It is especially interesting that in all three calculations, the peak occurs at $\approx 2.0 - 2.5$ fm, which is within the range of impact parameters where the gradient expansion that we use can be trusted. 
However, it is important to note that the values for angular momentum obtained in \cite{Gao:2007bc, Becattini:2007sr} are $\approx 10^5$ at RHIC energies, and  even larger at LHC energies, and are thus five to six orders of magnitude larger than our results. 

\begin{figure}[H]
\centering
\includegraphics[scale=0.97]{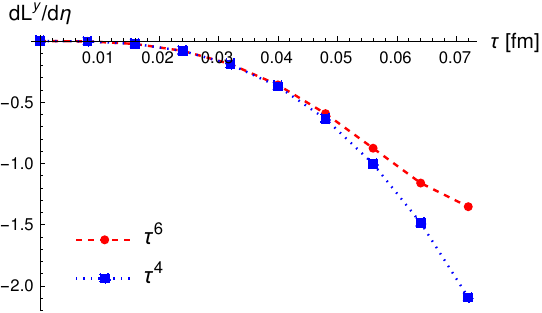} 
\caption{The angular momentum per unit rapidity  as a function of proper time with $b=2$ fm using $R_{\rm max} = 5.9$ fm. 
\label{plotLtime}}
\end{figure}

In figure \ref{plotLtime} we show the angular momentum as a function of proper time, at fixed impact parameter $b=2$ fm at different orders in the $\tau$ expansion. We have used the value $R_{\rm max}=5.9$ fm in all cases. The figure shows that the $\tau$ expansion appears to converge for much larger times than the other quantities we have calculated. 
The authors of Ref. \cite{Fries:2017ina} observed that the near field expansion method can produce very large values of angular momentum, if one considers large enough times. 
We have verified that our calculation reproduces this behaviour, but the sign of the angular momentum changes, and the numerical value of the result depends strongly on the order of the $\tau$ expansion. Both of these properties indicate that there is no reason to believe  the result is physical. 
 
Our results indicate that the glasma carries only a very small imprint of the primordial angular momentum, which means that the majority of the angular momentum is carried by valence quarks, for times at which the calculation is valid. This result casts doubt on the idea of 
a rapidly rotating initial glasma state that could be observed via polarization of final state hadrons. 

\section{Conclusions}
\label{sec-conclusions}

We have used a CGC approach and an expansion in proper time to derive an analytic result for the glasma energy-momentum tensor to sixth order in the proper time. 
We have taken into account some aspects of nuclear structure using spatially dependent nuclear density functions. 
In our previous paper \cite{Carrington:2020ssh} we gave a detailed description of the steps involved in our calculation. 
In this paper, we have concentrated on the physical results that can be obtained from our final expression for the energy-momentum tensor. 

For most of the quantities that we calculated, the proper time expansion can be trusted to about $\tau=0.05$ fm, and we have 
shown that the glasma moves towards equilibrium until the time at which the near field expansion breaks down.
Using simple arguments based on the uncertainty principle \cite{Carrington:2020ssh}, one can argue that this upper bound of the region where
the near field expansion converges reaches far beyond the lower bound at which we no longer
trust the classical description that is inherent in our method.
Our calculation also requires an expansion in  gradients of the nuclear density function. 
Some of the quantities we have calculated are almost entirely insensitive to this expansion, and in other cases the gradient expansion restricts us to the consideration of small impact parameters. 

We have calculated the first three Fourier coefficients of the azimuthal distribution of the momentum field of the glasma. 
Our results are larger than is generally expected, which is interesting because it contradicts the usual assumption that azimuthal anisotropy is mostly generated during the hydrodynamic evolution of the plasma. 
We have also calculated the eccentricity of the glasma, which describes the  spatial azimuthal asymmetry of the system. 
We have found a sizable correlation between the elliptic flow coefficient and the eccentricity, which indicates that the spatial asymmetry introduced by the initial geometry is effectively transmitted to the azimuthal distribution of the gluon momentum field. 
This result is interesting because a correlation of this kind is  an indication of the onset of hydrodynamics. 
We have formulated a calculation of the angular momentum per unit rapidity, on a hypersurface of constant proper time. 
Our results are much smaller than the total angular momentum of the participating nucleons \cite{Gao:2007bc,Selyuzhenkov_2011,Becattini:2007sr}, which shows that most of the angular momentum carried by valence quarks is not transmitted to the glasma. 
These results are significant because they contradict the picture of a rapidly rotating initial glasma state that has been proposed by several authors \cite{Liang:2004xn,Liang:2019clf} and led to experimental searches for a polarization effect in the hyperons and vector mesons produced in heavy ion collisions, which have been, so far, largely unsuccessful \cite{Adam:2018ivw,Acharya:2019ryw}. 

Finally, we comment that all of the calculation done in this paper
make use of  analytic solutions of the near field expanded YM equation which we have obtained to order $\tau^6$, and these results are also useful in other
contexts. 
We are currently using these solutions to perform a calculation of the transport properties of
heavy quarks in glasma. 

\section{Acknowledgments}

This work was partially supported by 
the Natural Sciences and Engineering Research Council of Canada from grant 2017-00028
and National Science Centre, Poland under grant 2018/29/B/ST2/00646.

\appendix

\section{Notation}
\label{app-notation}
%
The collision axis is defined to be the $z$-axis and the two transverse coordinates are denoted $\vec x_\perp$.
We use Minkowski, light-cone and Milne coordinates in different parts of the calculation, and these  coordinates are written $(t,z,\vec x_\perp)$, $(x^+,x^-,\vec x_\perp)$ and $(\tau,\eta,\vec x_\perp)$.  We use the conventional definitions
\bea
&& x^+=\frac{t+z}{\sqrt{2}} \text{~~and~~} x^-=\frac{t-z}{\sqrt{2}}\\
&& \tau = \sqrt{t^2-z^2}=\sqrt{2x^+ x^-} \text{~~and~~} \eta = \frac{1}{2} \ln \left( \frac{x^+}{x^-} \right)\,.
\eea
We define the relative and average transverse coordinates
\bea
&& \vec r = \vec x_\perp-\vec y_\perp \text{~~and~~}
 \vec R = \frac{1}{2}\left(\vec y_\perp+\vec x_\perp\right) \label{rR-def}\,.
\eea
We will write unit vectors as $\hat r=\vec r/|\vec r|=\vec r/r$ and $\hat R=\vec R/|\vec R|=\vec R/R$
and use standard notation for derivatives, for example
\bea
\partial^i_{x} = -\nabla_x^i =  -\frac{\partial}{\partial x_\perp^i} 
\eea
and similarly for $y_\perp$, $\vec r$ and $\vec R$. 

The metric tensors in these three coordinate systems are $g^{\rm mink} = (1,-1,-1,-1)_{\rm diag}$ and 
\bea
g_{\mu\nu}^{\rm lc} =
\left(
\begin{array}{cccc}
 0 & 1 & 0 & 0 \\
 1 & 0 & 0 & 0 \\
 0 & 0 & -1 & 0 \\
 0 & 0 & 0 & -1 \\
\end{array}
\right)\,,\qquad
g_{\mu\nu}^{\rm milne}=  \left(
\begin{array}{cccc}
 1 & 0 & 0 & 0 \\
 0 & -\tau^2 & 0 & 0 \\
 0 & 0 & -1 & 0 \\
 0 & 0 & 0 & -1 \\
\end{array}
\right)\,.\label{metric}
\eea
The coordinate transformations are
\bea
&& x^\mu_{\rm mink} = M^\mu_{~\nu} x^\nu_{\rm lc}\,,\qquad 
M^\mu_{~\nu} = \frac{dx^\mu_{\rm mink}}{dx^\nu_{\rm lc}} = 
\left(
\begin{array}{cccc}
 \frac{1}{\sqrt{2}} & \frac{1}{\sqrt{2}} & 0 & 0 \\
 \frac{1}{\sqrt{2}} & -\frac{1}{\sqrt{2}} & 0 & 0 \\
 0 & 0 & 1 & 0 \\
 0 & 0 & 0 & 1 \\
\end{array}
\right) \nn \\
&& x^\mu_{\rm mink} = M^\mu_{~\nu} x^\nu_{\rm milne}\,,\qquad 
M^\mu_{~\nu} = \frac{dx^\mu_{\rm mink}}{dx^\nu_{\rm milne}} = 
\left(
\begin{array}{cccc}
 \cosh (\eta ) & \tau  \sinh (\eta ) & 0 & 0 \\
 \sinh (\eta ) & \tau  \cosh (\eta ) & 0 & 0 \\
 0 & 0 & 1 & 0 \\
 0 & 0 & 0 & 1 \\
\end{array}\label{milne2mink}
\right)\,.
\eea

The generators $t_a$ of SU$(N_c)$ satisfy
\bea
\label{gen-def}
&& [t_a,t_b] = i f_{abc} t_c  \nonumber\\
&& \text{Tr}(t_a t_b) = \frac{1}{2}\delta_{ab} \nonumber \\
&& f_{abc} = -2i \text{Tr}\big(t_a[t_b,t_c]\big) \,.
\eea
Functions like $A_\mu$, $J_\mu$, $\rho$ and $\Lambda$ are SU$(N_c)$ 
valued functions and can be written as linear combinations of the SU$(N_c)$ generators.
In the adjoint representation we write the generators with a tilde as $(\tilde t_a)_{bc} = -i f_{abc}$. 

\section{Coefficients of the energy-momentum tensor at order $\tau^4$}
\label{sec-order4}

In this appendix we give the coefficients ${\cal X}_n^{lm}$ of equation (\ref{TTgeneric7}) with $n=4$. 
We remind the reader of the notation we are using. 
We have defined $L\equiv\ln(Q_s/m)$, 
$\hat\mu_1(\vec R) \equiv \mu_1(\vec R)/\bar\mu$, 
$\hat\mu_2(\vec R) \equiv \mu_2(\vec R)/\bar\mu$, 
$\bar\mu\equiv Q^2_s/g^4$, and we use the shorthand notation 
$\hat\mu_1^{10} \equiv \nabla_x \hat\mu_1(\vec R)$, $\hat\mu_1^{11} \equiv \nabla_x \nabla_y \hat\mu_1(\vec R)$, etc. Our fourth-order results are
\bea
\beta_4^{10}  &=&
 \frac{\hat\mu _1 Q_s^8}{65536 \pi ^4 g^2}
 \bigg(
-9 \hat\mu _2^2 (1-2 L)^4 \left(5828 \hat\mu _1{}^{10}-3449 \hat\mu _2{}^{10}\right)-46800 \pi  \hat\mu _2 (1-2 L)^2 \hat\mu
   _1{}^{10} \nn \\
&   +& \left(-31041 \hat\mu _1^2 (1-2 L)^4-48240 \pi  \hat\mu _1 (1-2 L)^2+5120 \pi ^2 (L+1)\right) \hat\mu _2{}^{10}
\bigg) - (\hat\mu_1\leftrightarrow \hat\mu_2) \nn \\
 \gamma_4^{11}  &=& -\frac{27 \hat\mu _1 Q_s^8}{32768 \pi ^4 g^2 m^2}
 \bigg(
(2 L-1) \left(\right.\hat\mu _2^2 (1-2 L)^2 \left(287 \hat\mu _1{}^{11}+216 \hat\mu _2{}^{11}\right) \nn\\
& +& \hat\mu _1 \left(175 \hat\mu _1 (1-2 L)^2+144 \pi
   \right) \hat\mu _2{}^{11}+216 \pi  \hat\mu _2 \hat\mu _1{}^{11}\left.\right)
\bigg)+(\hat\mu_1\leftrightarrow \hat\mu_2) \nn\\
\delta_4^{02}  &=& \frac{3 \hat\mu _1 Q_s^8}{65536 \pi ^4 g^2 m^2}
 \bigg(
9 \hat\mu _2^2 (2 L-1)^3 \left(2443 \hat\mu _1{}^{02}+2530 \hat\mu _2{}^{02}\right)+16080 \pi  \hat\mu _2 (2 L-1) \hat\mu _1{}^{02} \nn \\
& +&\left(9951 \hat\mu
   _1^2 (2 L-1)^3+8496 \pi  \hat\mu _1 (2 L-1)+256 \pi ^2\right) \hat\mu _2{}^{02}
\bigg) +(\hat\mu_1\leftrightarrow \hat\mu_2) \nn \\
\delta_4^{20}  &=& \frac{3 \hat\mu _1 Q_s^8}{65536 \pi ^4 g^2 m^2}
 \bigg(
9 \hat\mu _2^2 (2 L-1)^3 \left(1869 \hat\mu _1{}^{20}+2098 \hat\mu _2{}^{20}\right)+12192 \pi  \hat\mu _2 (2 L-1) \hat\mu _1{}^{20} \nn\\
& + & \left(6801 \hat\mu
   _1^2 (2 L-1)^3+5904 \pi  \hat\mu _1 (2 L-1)+256 \pi ^2\right) \hat\mu _2{}^{20}
\bigg)+(\hat\mu_1\leftrightarrow \hat\mu_2) \nn \\
 {\cal E}_4^{00}  &=& \frac{3 \hat\mu _1 \hat\mu _2 Q_s^8}{65536 \pi ^4 g^2}
 \bigg(
8376 \hat\mu _2^2 (1-2 L)^4 + 3 \hat\mu _2 \left(1797 \hat\mu _1 (1-2 L)^2+3904 \pi \right)  \left(2 L -1\right){}^2 \nn \\
& +& 256 \pi^2 (2 L+3)  
\bigg)+(\hat\mu_1\leftrightarrow \hat\mu_2) \nn \\
{\cal E}_4^{02}  &=& \frac{\hat\mu _1 Q_s^8}{32768 \pi ^4 g^2 m^2} 
 \bigg(
9 \hat\mu _2^2 (2 L-1)^3 \left(1078 \hat\mu _1{}^{02}+1157 \hat\mu _2{}^{02}\right)+7068 \pi  \hat\mu _2 (2 L-1) \hat\mu _1{}^{02} \nn \\
& + & 4 \left(1047
   \hat\mu _1^2 (2 L-1)^3+900 \pi  \hat\mu _1 (2 L-1)+32 \pi ^2\right) \hat\mu _2{}^{02}
   \bigg)+(\hat\mu_1\leftrightarrow \hat\mu_2)\,.\nn\\
\label{results_order4}
\eea

\section{Azimuthal distribution}
\label{sec-azi-distro}

In this section we define the azimuthal distribution of the flow vector $T^{i0}(\vec x_\perp)$.
The azimuthal angle is measured with respect to the  $x$-axis and is written
\bea
\label{phi-x-perp}
\varphi(\vec x_\perp) = \tan^{-1}\Big(\frac{T^{0y}(\vec x_\perp)}{T^{0x}(\vec x_\perp)}\Big)  
= \cos^{-1}\bigg(\frac{T^{0x}(\vec x_\perp)}
{\sqrt{\big(T^{0x}(\vec x_\perp)\big)^2 + \big(T^{0y}(\vec x_\perp)\big)^2 }}\bigg)\,.
\eea
We define the distribution 
\bea
\label{phi-distribution}
P(\phi) \equiv \frac{1}{\Omega}  \int d^2 \vec x_\perp \,\delta \big( \phi -\varphi(\vec x_\perp) \big) \,
W(\vec x_\perp) 
\eea
where we have introduced the weighting function
\bea
 W(\vec x_\perp)  \equiv \sqrt{\big(T^{0x}(\vec x_\perp)\big)^2 + \big(T^{0y}(\vec x_\perp)\big)^2} 
 \eea
and the normalization factor
 \bea
 \Omega \equiv \int d^2 \vec x_\perp \, W(\vec x_\perp) \,.
\eea
The distribution $P(\phi)$ can be decomposed into Fourier harmonics as
\bea
P(\phi) = \frac{1}{2\pi} \Big(1 +  2\sum_{n=1}^\infty v_n \cos(n \phi ) \Big) 
\eea
where coefficients $v_n$ are given by the relation
\bea
\label{vn-1}
v_n = \int_0^{2\pi} d\phi \, \cos(n\phi) \, P(\phi) \, . 
\eea

\end{document}